\documentclass[useAMS,usenatbib]{mngen}
\pdfoutput=1
\usepackage{natbib}
\usepackage{verbatim} 
\usepackage{amsmath} 
\usepackage{amsbsy}
\usepackage{amssymb}
\usepackage{mathrsfs} 
\usepackage{pdflscape} 
\usepackage{epstopdf}
\usepackage{graphicx}
\usepackage{deluxetable}
\usepackage{longtable}
\usepackage{multirow}

\title[The Pulsar Contribution to the Gamma-Ray Background]{The Pulsar Contribution to the Gamma-Ray Background}
\author[Claude-Andr\'e Faucher-Gigu\`ere and Abraham Loeb]{Claude-Andr\'e Faucher-Gigu\`ere$^{1}$\thanks{E-mail:
cgiguere@cfa.harvard.edu} and Abraham Loeb$^{1}$\thanks{E-mail: aloeb@cfa.harvard.edu}\\
$^{1}$Department of Astronomy, Harvard University, Cambridge, MA 02138, USA}
\begin{document}

\pagerange{\pageref{firstpage}--\pageref{lastpage}} \pubyear{2002}

\maketitle

\label{firstpage}

\begin{abstract}
We estimate the contribution of Galactic pulsars, both ordinary and millisecond pulsars (MSPs), to the high-energy ($>$100 MeV) $\gamma-$ray background.
We pay particular attention to the high-latitude part of the background that could be confused with an extragalactic component in existing analyses that subtract a Galactic cosmic-ray model.
Our pulsar population models are calibrated to the results of large-scale radio surveys and we employ a simple empirical $\gamma-$ray luminosity calibration to the spin-down rate that provides a good fit to existing data.
We find that while ordinary pulsars are expected to contribute only a fraction $\sim10^{-3}$ of the high-latitude $\gamma-$ray intensity ($I_{X}\sim1\times10^{-5}$ ph s$^{-1}$ cm$^{-2}$ sr$^{-1}$), MSPs could provide a much larger contribution and even potentially overproduce it, depending on the model parameters.
We explore these dependences using a range of MSP models as a guide to how $\gamma-$ray measurements can usefully constrain the MSP population.
Existing $\gamma-$ray background measurements and source counts already rule out several models.
Finally, we show how fluctuations in the $\gamma-$ray sky can be used to distinguish between different sources of the background.
\end{abstract}


\section{Introduction}
Identifying the nature of diffuse radiation backgrounds is a fundamental problem because it may both lead to the discovery of previously unknown astrophysical sources and constrain the properties of known ones.
While we now know the ultraviolet background to be dominated by quasars and star-forming galaxies 
\citep[e.g.,][]{2005MNRAS.357.1178B, 2008ApJ...682L...9F, 2008ApJ...688...85F, spectrum} and that the soft X-ray background consists chiefly  of emission from active galactic nuclei (AGN) \citep[much of which has now been resolved by the \emph{Chandra X-ray Observatory} and \emph{XMM Newton}; e.g.,][]{2005ARA&A..43..827B, 2006ApJ...645...95H, 2007ApJ...661L.117H}, the sources of the diffuse $\gamma-$ray background have yet to be convincingly established.
With the \emph{Fermi Gamma Ray Space Telescope}\footnote{http://fermi.gsfc.nasa.gov} (\emph{Fermi}) now taking data, it is timely to revisit this question from a theoretical point of view and make predictions that can be directly tested within the next few years.
At the same time, the \emph{AGILE}\footnote{http://agile.rm.iasf.cnr.it} telescope of the Italian Space Agency is making interesting discoveries in spite of its more limited capabilities in the $\gamma-$ray range \citep[e.g.,][]{2008ApJ...688L..33H, 2009arXiv0903.0087P}.
High-energy ($>100$ MeV) $\gamma-$rays are of particular interest as they probe some of the most energetic objects in the universe, involving sources such as pulsars \citep[e.g.,][]{1999ApJ...516..297T} and blazars \citep[e.g.,][]{1997ApJ...490..116M} emitting most of their electromagnetic luminosity in this band.\\ \\
In the standard picture, the $\gamma-$ray background is separated into a strong Galactic component and a nearly isotropic component believed to be of extragalactic origin.
The Galactic component is generally thought to be predominantly produced by cosmic-ray interactions with interstellar gas and radiation \citep[e.g.,][]{1993ApJ...416..587B, 1997ApJ...481..205H, 2000ApJ...537..763S}, while the isotropic component is commonly attributed to blazars \citep[e.g.,][]{1996ApJ...464..600S, 1998ApJ...494..523S}.
Some calculations however indicate that blazars may only account for a fraction of the total isotropic component \citep[e.g.,][]{1998ApJ...496..752C, 1999APh....11..213M, 2000MNRAS.312..177M} and all blazar estimates presently suffer from significant uncertainties.
For instance, the  blazars seen by the \emph{Energetic Gamma Ray Experiment Telescope}\footnote{http://heasarc.gsfc.nasa.gov/docs/cgro/egret} (\emph{EGRET}) on the \emph{Compton Gamma Ray Observatory} were almost always in a flaring state, with the duty cycle for flaring being uncertain, and the evolution of blazar $\gamma-$ray emission must be extrapolated beyond the existing data \cite[][]{2008RPPh...71k6901T}.
\emph{Fermi} has to date resolved only about 7\% of the isotropic component into blazars \citep[][]{2009ApJ...700..597A}.\\ \\
The separation of the $\gamma-$ray background into Galactic and isotropic components is a difficult task that relies on an accurate model for the Galactic contribution \citep[e.g.,][]{1998ApJ...494..523S, 2004JCAP...04..006K, 2004ApJ...613..956S}.
It is therefore important to investigate Galactic sources that may contribute significantly to the high-latitude background and be confused with the extragalactic background.
The detection of six pulsars at high confidence by \emph{EGRET} \citep[][]{1999ApJ...516..297T}, the recent discovery of a radio-quiet $\gamma-$ray pulsar at the center of the supernova remnant CTA 1 \citep[][]{2008Sci...322.1218A}, and the firm detection of at least 29 pulsars in the $\gamma-$rays by \emph{Fermi} \citep[][]{2009ApJS..183...46A} raise the possibility that pulsars could contribute a significant fraction of the high-latitude background.
The high space velocities of pulsars \cite[e.g.,][]{1970ApJ...160..979G, 1994Natur.369..127L, 1997MNRAS.291..569H, 2002ApJ...568..289A, 2006ApJ...643..332F} imply that their spatial distribution may be extended above and below the Galactic plane and not correlate well with other tracers used to separate the Galactic component.\\ \\
Another timely motivation for understanding the sources of the $\gamma-$ray background is provided by indirect searches for dark matter annihilation products, which may be detectable in $\gamma-$rays \citep[e.g.,][]{2007ApJ...657..262D, 2008ApJ...686..262K, 2008JCAP...07..013B, 2008Natur.456...73S, 2008PhRvL.101z1301K, 2009arXiv0903.2829D}.
Among the possible astrophysical sources, $\gamma-$ray pulsars might be the most difficult to disentangle from a dark matter annihilation signal owing to their similar spectra \citep[][]{2007ApJ...659L.125B}.
Given the low intensity expected from dark matter annihilation, even a small contribution by pulsars to the $\gamma-$ray sky could be an important contaminant for these searches.\\ \\
The idea of pulsars as contributors to the $\gamma-$ray background has previously received some attention in the literature.
Ordinary pulsars have been suggested to provide a significant fraction of the low-latitude Galactic $\gamma-$ray background, with estimates ranging from 5-10\% to comparable to the cosmic ray contribution \citep[][]{1976ApJ...208L.107H, 1977MNRAS.179P..69S, 1981Natur.290..316H, 1981ApJ...247..639H, 1991JApA...12...17B, 1996ApJ...461..872S, 2000ApJ...538..818M}.
It has also been proposed that millisecond pulsars (MSPs), owing to their rapid spin frequencies, may be bright in the $\gamma-$rays \citep[][]{1983Natur.305..409U, 1994A&A...281L.101S, 2005ApJ...622..531H} and therefore also contribute significantly to the $\gamma-$ray background \citep[][]{1991JApA...12...17B, 1992ApJ...391..659B, 1997ApJ...476..238B}.\\ \\
The present work aims to improve previous crude estimates of the pulsar contribution to the high-energy $\gamma-$ray background by using the pulsar population synthesis code developed by \cite{2006ApJ...643..332F} (FGK06), with the goals of making predictions for upcoming \emph{Fermi} observations and providing a guide to use these to constrain the population and astrophysics of Galactic pulsars.
The population synthesis code allows us to consider pulsar populations with realistic distributions of physical properties and calibrated against existing large-scale radio surveys.
As we will argue, however, simple physical arguments and comparison with existing $\gamma-$ray data also allow relatively model-independent conclusions to be drawn.
We pay particular attention to the pulsar contribution to the high-latitude background that may be confused with the extragalactic component or the dark matter signal. 
A similar approach was used by \cite{2007Ap&SS.309...35S} to quantify the contribution of unresolved sources to the Galactic diffuse emission.\\ \\
We begin by describing the population synthesis models used to calculate the contribution of pulsars to the $\gamma-$ray background in \S \ref{population synthesis models}.
Our results are compared with existing $\gamma-$ray observations in \S \ref{comparison with existing gamma rays} and the implications for pulsar physics are discussed \S \ref{physics implications}.
In \S \ref{pixel fluctuations}, we show how fluctuations in the $\gamma-$ray sky can be used to discriminate between different origins of the $\gamma-$ray background.
We finally conclude in \S \ref{conclusion}.

\section{Population Synthesis Models}
\label{population synthesis models}
We first estimate the expected pulsar contribution to the $\gamma-$ray sky by combining pulsar population synthesis models calibrated against radio surveys with a prescription motivated by existing $\gamma-$ray observations for the $\gamma-$ray luminosity of pulsars.

\subsection{Models}
\label{population synthesis}
The basis for our calculations of the $\gamma-$ray emission from pulsars is the population synthesis code developed by FGK06, who studied the birth properties and evolution of isolated radio pulsars.
In that work, a Milky Way model was seeded with nascent pulsars which were physically evolved in time.
The synthetic pulsar population was then observed by mock realizations of the Parkes and Swinburne Multibeam Pulsar Surveys \citep[][]{2001MNRAS.328...17M, 2001MNRAS.326..358E}, including all the main selection effects, and the results compared with the actual detections.
This resulted in a model for the spatial, kinematic, rotational, magnetic, and radio properties of the pulsar population that reproduces the observations well.\\ \\
In this work, we use this model to calculate the $\gamma-$ray contribution of ordinary pulsars.
As will be discussed below, we extend the code to also model a population of MSPs.
The evolution of these recycled pulsars, whose spin-up to millisecond periods is thought to have occurred by angular momentum accretion from a binary companion, is much more complex \citep[e.g.,][]{1991PhR...203....1B}.
We therefore do not attempt to model their evolution but rather prescribe probability distributions for the properties of their current Galactic population.
This population is also not as well constrained observationally, both due to the relatively small number of known MSPs and to the difficulty of accurately modeling the selection effects that affect their detection.
For example, the scattering of radio waves by free electron density fluctuations \citep[e.g.,][]{1990ARA&A..28..561R} makes it challenging to detect short pulsar periods toward the Galactic center and the orbital acceleration in binary systems acts to mask periodicity \citep[e.g.,][]{1991ApJ...368..504J}.
Another potentially important selection effect against the detection of MSPs is that obscuration of the pulsar signal by material blown off a companion by the pulsar wind may render it invisible, especially in the radio \citep[][]{1991ApJ...379L..69T}.
The fastest-spinning known MSP J1748$-$2446ad is in fact eclipsed in the radio for $\sim40$\% of its orbit \citep[][]{2006Sci...311.1901H}.\\ \\
As will be argued (see \S \ref{ordinary vs msp} and \S \ref{comparison with existing gamma rays}), however, simple physical arguments can be used to draw conclusions on the contribution of MSPs to the high-latitude $\gamma-$ray background that are insensitive to the details of the population models (see also Strong 2007\nocite{2007Ap&SS.309...35S} for a similar approach in the context of the Galactic diffuse emission).

\subsection{Calibration Against Radio Surveys}
\label{radio calibration}
Most of the constraints on pulsar populations come from radio observations, with about 1800 objects now documented in the ATNF Pulsar Database.\footnote{http://www.atnf.csiro.au/research/pulsar/psrcat/ \citep[][]{2005AJ....129.1993M}.}
Since we will explore models beyond the ones studied by FGK06 for MSPs, we pause to summarize how we model the sensitivity of radio surveys to pulsars.
The mock surveys will be used to calibrate our models based on the actual observations.\\ \\
The detectability of a pulsar depends on its intrinsic properties (luminosity, pulse period, and duty cycle), on its location (distance, dispersion measure, interstellar scattering, and brightness temperature of the background sky), and on the details of the observing system.
The minimum flux theoretically detectable from a radio pulsar is estimated by the usual radiometer equation,
\begin{equation}
\label{radiometer equation}
S_{\rm min}=\delta_{\rm beam} \frac{\beta_{\rm sys} \sigma (T_{\rm rec}+T_{\rm sky})}{G\sqrt{N_{p}\Delta \nu t_{\rm int}}}\sqrt{\frac{W_{e}}{P-W_{e}}},
\end{equation}
where $\delta_{\rm beam}$ is a factor accounting for the reduction in sensitivity to pulsars located away from the center of the telescope beam, $T_{\rm rec}$ is the receiver temperature on a cold sky, $T_{\rm sky}$ is the sky background temperature, $G$ is the antenna gain, $N_{p}$ is the number of polarizations summed, $\Delta \nu$ is the receiver bandwidth, $t_{\rm int}$ is the integration time, $P$ is the pulse period, $W_{e}$ is the effective pulse width, $\sigma$ is the signal-to-noise detection threshold, and $\beta_{\rm sys}$ is a constant accounting for various system losses \citep[e.g.,][]{dss+84}.\\ \\
Further details regarding the $S_{\rm min}$ calculation are provided in Appendix 1 of FGK06.
Briefly, the effective pulse width $W_{e}$ is a quadratic sum of the pulsar's intrinsic pulse width, the effective sampling time of the survey, the dispersion smearing across a receiver channel, and the pulse scattering time scale.
The dispersion measure and pulse scattering time of each simulated pulsar are calculated using the NE2001 model for the Galactic free electron density distribution \citep[][]{NE2001} given the sightline through the Galaxy.
The sky brightness temperature at 408 MHz is taken from the \cite{1981A&A...100..209H} all-sky map and scaled to the desired frequency assuming a spectral index $\alpha_{\rm bg}=2.7$ \citep[$T_{\rm sky}({\nu})\propto \nu^{-\alpha_{\rm bg}}$;][]{1987MNRAS.225..307L}.
The $\delta_{\rm beam}$ flux degradation factor is calculated assuming a Gaussian power pattern, with each synthetic pulsar in the covered sky area lying at a random position within the half-power cone of a telescope beam.
A synthetic pulsar is detected if and only if it lies in the sky area covered by a survey, the radio flux density of the pulsar at the central frequency of the survey at the Earth exceeds the survey threshold given by equation (\ref{radiometer equation}), and the pulsar is actually beamed toward the observer in the radio.
The latter criterion is implemented by assigning a probability $f_{\rm b}^{\rm radio}$ (which in general can depend on the period of the object) to each pulsar that its beam crosses our line of sight to it.\\ \\
The radiometer equation (\ref{radiometer equation}) gives the minimum flux for the detection of a pulsed signal and is valid for solitary pulsars.
While less than 1\% of ordinary pulsars are observed to have orbiting companions, about 80\% of MSPs are found in binary systems \citep[e.g.,][]{2005LRR.....8....7L}.
The periods of pulsars in binary systems are modulated by the Doppler effect induced by orbital acceleration.
This acceleration tends to mask the periodicity of the signal by spreading it into multiple Fourier bins \citep[e.g.,][]{1991ApJ...368..504J}.
The resulting degradation in sensitivity is a function of the orbital parameters of the binary system and is in general most important when the orbital period is less than the integration time.\footnote{For integration times that are much less than the orbital period, the Doppler effect is nearly constant during the observation and the periodicity is approximately preserved.}
The most disfavored pulsars are therefore MSPs with close and/or massive companions.
As a result, it is difficult to quantify the number and characteristics of the binary pulsars that are missed in existing surveys.
Most, if not all, MSP population studies to date have neglected these effects.
In this work, we simply note that the unmodelled degradation in sensitivity to radio MSPs in binary systems can only lead us to underestimate the number of MSPs in the Galaxy (by optimistically estimating their detection probability) and thus to underestimate their cumulative $\gamma-$ray emission in particular models.

\subsection{Gamma-Ray Luminosity Prescription}
\label{gammaray luminosity}
In order to calculate the pulsar contribution to the $\gamma-$ray background, we must assign a $\gamma-$ray luminosity $L_{\gamma}$ to each pulsar in our radio-calibrated populations.
No definitive prescription for doing so exists at present and \emph{Fermi} is poised to provide unprecedented observational constraints on theoretical models.
Our approach is to adopt an empirically-calibrated prescription, which we will use to make predictions that can be tested with upcoming observations.\\ \\
Unlike in the radio, where the emitted power is a negligible fraction of the total emission, a significant fraction of the rate of loss of rotational kinetic energy, $\dot{E}$, of pulsars is observed in the form of $\gamma-$rays.
In fact, the maximum observed energy output is in the $\gamma-$ray band for all the pulsars detected by \emph{EGRET}.
For these sources, the ratio of the electromagnetic power above a photon energy of 1 eV to the total spin-down power ranges between about 0.1\% and 20\%, assuming a $\gamma-$ray beam covering 1 steradian \citep[][]{1998MNRAS.295..337R, 1999ApJ...516..297T}.
For a flat spectrum, with equal power per energy decade, half of this power is in the form of $>100$ MeV $\gamma-$rays.
The flux measured by \emph{Fermi} for the pulsar at the center of the CTA 1 supernova remnant also indicates a $\gamma-$ray efficiency of about 1\% for the same beaming solid angle, but perhaps as high as 10\% if the emission arises in the outer magnetosphere, for which models predict beam sizes $\gg1$ sr \citep[][]{2008Sci...322.1218A}.\\ \\
The few pulsars studied in the $\gamma-$rays by \emph{EGRET} suggest a simple relation
\begin{equation}
\label{sqrt Edot relation}
L_{\gamma} \propto \sqrt{\dot{E}}\propto \dot{P}^{1/2} P^{-3/2},
\end{equation}
where $\dot{E}=4 \pi^{2} I \dot{P}/P^{3}$ is the rate of loss of rotational kinetic energy of the neutron star, with $I$ (conventionally assumed equal to $10^{45}$ g cm$^{2}$) denoting its moment of inertia.
This relation was first noticed and theoretically investigated by \cite{1996A&AS..120C..49A} \citep[see also][]{1999ApJ...516..297T}.
No other simple parameters appear well correlated with the observed properties of these pulsars \citep[][]{1996A&AS..120C.103G}.
In particular, there appears to be more variance in the fraction of $\dot{E}$ that is converted into $\gamma-$rays.\\ \\
The empirical relation in equation (\ref{sqrt Edot relation}) can also be motivated theoretically by the facts that the total voltage drop in the magnetosphere $\Delta V\propto \sqrt{\dot{E}}$ \citep[][]{1975ApJ...196...51R, 1996A&AS..120C..49A} or that the \cite{1969ApJ...157..869G} current, defined as the rate of outflowing relativistic particles over the polar cap, $\dot{N}_{\rm GJ}\propto \sqrt{\dot{E}}$ as well \citep[][]{1981ApJ...245..267H}.
In outer gap models of high-energy emission, this relation can also be approximately derived as the product of the volume of the outer gap, the particle number density assumed to be close to the \cite{1969ApJ...157..869G} value, and the power radiated by curvature radiation by a single particle \citep[e.g.,][]{1996ApJ...470..469R, 2002nsps.conf..162C}.\\ \\
To avoid spectral uncertainties in the conversion from energy units to photon units, we express luminosities directly in units of ph s$^{-1}$ above $100$ MeV, and denote them $L_{\gamma}^{\rm ph}$.
Since the $\gamma-$ray fluxes reported by \emph{Fermi} are already expressed in ph s$^{-1}$ cm$^{2}$, our empirical calibration (\S \ref{observed gamma ray luminosities}) is free of any assumption with regards to the slope or high-energy cutoff of the $\gamma-$ray pulsar spectrum.
 More precisely, we define the $\gamma-$ray photon luminosity of a pulsar above 100 MeV to be given by
\begin{equation}
\label{sqrt Edot prescription}
L_{\gamma}^{\rm ph} \equiv K \min\{C \dot{P}^{1/2} P^{-3/2},~f_{\gamma}^{\rm max} \dot{E}\},
\end{equation}
where $C$ is an empirically-determined proportionality constant, $f_{\gamma}^{\rm max}$ is the maximum fraction of the rotational power loss converted into $\gamma-$rays, and $K$ accounts for the conversion from erg s$^{-1}$ to ph s$^{-1}$.

\begin{figure}
\includegraphics[width=0.475\textwidth]{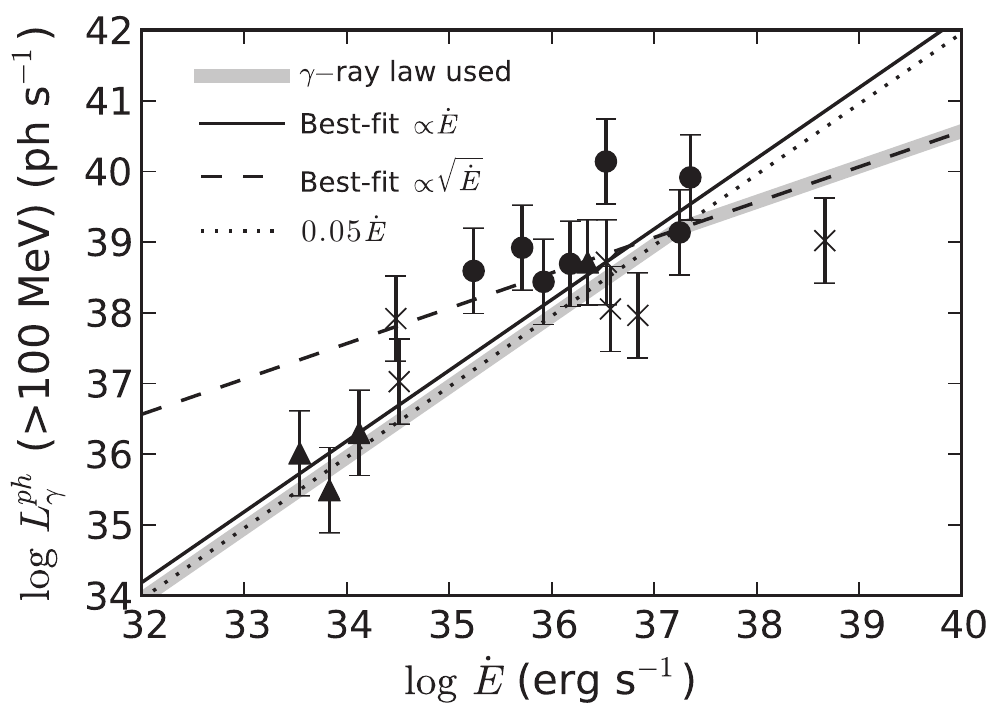}
\caption[]{Estimated $\gamma-$ray photon luminosity at energies $>$100 MeV vs. rate of loss of rotational kinetic energy for the firm pulsar detections in the first three months of \emph{Fermi} data and in the first year of \emph{AGILE} data for which rotational parameters and distances are tabulated in the ATNF database.
The crosses show the \emph{EGRET} pulsars (from left to right: B1055$-$52, B1951+32, B1706$-$44, J0633+1746, B0833$-$45, B0531+21).
The triangles indicate MSPs (J0030+0451, J2124$-$3358, J0613$-$0200, B1821$-$24*).
The filled circles show other ordinary pulsars detected by \emph{Fermi} or \emph{AGILE} (J0631+1036, J1509$-$5850, J1028$-$5819, B1046$-$58, J2021+3651, B1509$-$58*, J2229+6114*).
The asterisks indicate \emph{AGILE} discoveries.
The point for B1046$-$58 was translated to the left by $\Delta \dot{E}=-5\times10^{-35}$ erg s$^{-1}$ to distinguish it from MSP B1821$-$24. 
The solid line shows the best-fit $L_{\gamma}^{\rm ph}\propto\dot{E}$ law and the dashed line the best-fit $L_{\gamma}^{\rm ph}\propto\sqrt{\dot{E}}$ law.
The MSP points were excluded for these fits.
The dotted line indicates $L_{\gamma}=0.05\dot{E}$ under the assumption of a spectrum with equal power per decade energy and a high-energy cutoff $E_{\rm max}=3$ GeV.
The $\gamma-$ray luminosity law adopted in our calculations is shown by the broken thick grey line. 
The calculated $\gamma-$ray luminosities assume isotropic emission.}
\label{Lgamma vs Edot}
\end{figure} 

\subsection{Observed Gamma-Ray Luminosities}
\label{observed gamma ray luminosities}
The parameters of our $\gamma-$ray luminosity prescription are set based on the measured properties of the known $\gamma-$ray pulsars.
Specifically, we consider the subset of the 29 firm pulsar detections in the first three months of \emph{Fermi} data \citep[][]{2009ApJS..183...46A} whose rotational parameters and distances are tabulated in the ATNF database and the three firm first-year $\gamma-$ray discoveries by \emph{AGILE} \citep[two of these are not part of the firm \emph{Fermi} detections;][]{2009arXiv0903.0087P}.
This excludes 14 objects newly discovered by \emph{Fermi} and for which these parameters (usually derived from radio counterparts, which may not exist in these cases) are not yet available.
This however includes all the \emph{EGRET} pulsars and three $\gamma-$ray MSPs.
Figure \ref{Lgamma vs Edot} shows the $\gamma-$ray photon luminosity for energies above 100 MeV for these pulsars versus their inferred spin-down luminosity.
The $\gamma-$ray photon luminosity was in each case estimated as $L_{\gamma}^{\rm ph}=4\pi d^{2} F_{\gamma}^{\rm ph}$, where $F_{\gamma}^{\rm ph}=F_{23} + F_{35}$, and $F_{23}$ and $F_{35}$ are the photon fluxes for energies 100 MeV$-$1 GeV and 1 GeV$-$100 GeV for each pulsar given by \cite{2009ApJS..183...46A}.
For the \emph{AGILE} detections with no \emph{Fermi} counterpart in \cite{2009ApJS..183...46A}, the photon fluxes at $>$100 MeV were taken from \cite{2009arXiv0903.0087P}.
The spin-down luminosities $\dot{E}$ and best distance estimates $d$ were taken from the ATNF database.
One exception is for PSR J2021+3651, for which we assumed a distance 12 kpc quoted by \cite{2004ApJ...612..389H} and consistent with the NE2001 free electron density model.
The error bars account for a factor of 2 uncertainty in $d$, roughly what can be conservatively expected for distances determined from the dispersion measure.
Variance in the viewing geometry may introduce additional scatter.
The uncertainty in the $\gamma-$ray flux is generally subdominant.\\ \\
Figure \ref{Lgamma vs Edot} also shows best-fit $L_{\gamma}^{\rm ph}\propto\dot{E}$ and $L_{\gamma}^{\rm ph}\propto\sqrt{\dot{E}}$ lines.
For these fits, the MSPs were excluded as a test of how well the fits to the better-studied ordinary pulsar data predict the MSP $\gamma-$ray luminosities. 
The prescription $L_{\gamma}^{\rm ph}\propto \sqrt{\dot{E}}$ suggested by the \emph{EGRET} pulsars holds reasonably well for this extended data set if we take into account the intrinsic scatter that may exist in the true $\gamma-$ray luminosity law and the uncertainties in the distances and beaming angles of the pulsars.
The extrapolation to the MSPs is excellent, in agreement with \cite{2009Sci...325..848A}, who find that the basic emission mechanism seems to be the same for MSPs and young pulsars. 
Again, because the empirical luminosities are already expressed in ph s$^{-1}$, the resulting best-fit relations for $L_{\gamma}^{\rm ph}=L_{\gamma}^{\rm ph}(\dot{E})$ do not depend on the spectral parameters.
While this has no impact at all on our other calculations, we may assume that pulsars have spectra containing approximately equal power per energy decade up to a high-energy cutoff $E_{\rm max}\approx3$ GeV to convert the photon luminosities to cgs units.
The dotted line in Fig. \ref{Lgamma vs Edot} indicates $L_{\gamma}=0.05\dot{E}$ for this fiducial choice and shows that
physically limiting the $\gamma-$ray luminosity at $0.05\dot{E}$ ($f_{\gamma}^{\rm max}=0.05$) provides an excellent match to the MSPs.
The resulting $\gamma-$ray luminosity law, used in the rest of this work, is shown by the broken thick grey line.\\ \\
After our analysis was completed, the first \emph{Fermi} $\gamma-$ray pulsar catalog, containing more pulsars than we have analyzed, was released \citep[][]{2009arXiv0910.1608A}. 
We have verified that our best-fit  $L_{\gamma}^{\rm ph}=L_{\gamma}^{\rm ph}(\dot{E})$ relation is broadly consistent with the expanded data set. 
Moreover, as our main results are only weakly dependent on the details of the population synthesis assumptions by virtue of their calibration to existing source counts (see \S \ref{ordinary vs msp} and \S \ref{comparison with existing gamma rays}), they are unaffected by the new data. 

\subsection{Gamma-Ray Beaming Fraction}
\label{gamma ray beaming fraction}
The $\gamma-$ray luminosities above were derived assuming a $\gamma-$ray beam size of 4$\pi$ sr.
This assumption has no effect on the $\gamma-$ray flux observed at the Earth for any given pulsar; while the inferred $\gamma-$ray luminosity of the pulsar scales with the beam size, so does the factor by which the flux is diluted as it propagates toward us.
However, larger beam sizes imply that more pulsars have their beams intersecting our sightline to them and are actually observable from the Earth.
This effect therefore enhances the predicted $\gamma-$ray background for particular models.
Because the shape of the flux distribution is determined predominantly by the geometry of the spatial distribution (see \S \ref{comparison with existing gamma rays}), models which predict the same number of detections above a given flux threshold also however predict approximately the same integrated $\gamma-$ray background.\\ \\
The beam sizes of pulsars determine the fraction $f_{\rm b}$ of all pulsars whose beams cross our sightline. 
Half of the firm pulsar detections in the first three months of \emph{Fermi} data are new discoveries with no previously known radio counterparts.
Since it is in principle easier to detect and confirm $\gamma-$ray pulsars associated with known sources, this suggests that the $\gamma-$ray beaming fraction $f_{\rm b}^{\gamma}$ is in general much larger the radio beaming fraction $f_{\rm b}^{\rm radio}$, and likely of order unity.
Theoretical models in which the $\gamma-$ray emission originates from the outer or slot gaps of pulsar magnetospheres in fact predict large $\gamma-$ray beams of size $\gg1$ sr \citep[e.g.,][]{1995ApJ...438..314R, 2008ApJ...688L..25H, 2008ApJ...680.1378H}.
Comparatively large $\gamma-$ray beaming fractions would also be consistent with many of the \emph{EGRET} unidentified sources being pulsars in spite of the mostly unsuccessful searches for radio pulsars within their error boxes \citep[e.g.,][]{2003MNRAS.342.1299K, 2005MNRAS.364.1011C, 2006ApJ...652.1499C}.
Note that in the limit of large $\gamma-$ray beams, the details of the beaming geometry have no impact on predictions of the integrated $\gamma-$ray background.\\ \\
The beaming fractions of ordinary and millisecond pulsars in the radio, which are important in normalizing their populations to radio survey detections, will be treated separately in \S \ref{ordinary pulsars} and \S \ref{msps}.

\subsection{Ordinary Pulsars}
\label{ordinary pulsars}
For the ordinary pulsars, the population parameters and number normalization were determined in detail by FGK06 through a comparison with the 1.4 GHz Parkes and Swinburne Multibeam Pulsar Surveys \citep[][]{2001MNRAS.328...17M, 2001MNRAS.326..358E}.
These two surveys, which together detected over 1000 of the 1800 pulsars in the ATNF database and which were performed with the same observing system, provide a comparison sample with a homogeneous selection function.
The ordinary pulsar population parameters are therefore taken as fixed.\\ \\
Briefly, the synthetic population of ordinary pulsars is an evolved snapshot of pulsars that are born in the Galactic spiral arms with a Galactocentric radial distribution peaking at radius $\approx3$ kpc and with a mean birth velocity of 380 km s$^{-1}$.
The birth spin period distribution is normal, with $\langle P_{0} \rangle=300$ ms and $\sigma_{P_{0}}=150$ ms, truncated so that $P_{0}>0$.
The magnetic field distribution is lognormal, with $\langle \log B\rangle=12.65$ and $\sigma_{\log B}=0.55$.
Throughout, magnetic fields are expressed in G and $\log$ refers to the base-10 logarithm.
The pulsars are evolved in time by solving for their orbits in the Galactic potential and by assuming magnetic dipole breaking with constant magnetic fields.
The radio luminosities of pulsars are also assumed to be proportional to $\sqrt{\dot{E}}$, as was found to produce a good fit to the $P-\dot{P}$ diagram, but with a random dithering factor.
The period-dependent radio beaming fraction is 
the one obtained by \cite{1998MNRAS.298..625T} in their analysis of polarization data:
\begin{equation}
\label{tauris manchester fb}
f_{\rm b}^{\rm radio,ord}(P) = 0.09 [\log(P/{\rm s}) - 1]^{2} + 0.03.
\end{equation}

\subsection{Millisecond Pulsars}
\label{msps}
Our population synthesis treatment for MSPs differs from our treatment of ordinary pulsars in that we do not attempt to model their time evolution.
Rather, we populate a present-day synthetic Galaxy with MSPs whose properties are drawn from prescribed distributions.
Since the constraints on this population are poorer, due to their more complex evolutionary channels and more severe selection effects, we will explore a range of models.
This approach will also serve as a guide of how $\gamma-$ray observations can provide new constraints on MSPs.
In this work, we do not attempt to constrain the details of the MSP population with radio data as was done for ordinary pulsars in FGK06 \citep[for work in this direction, see][]{1997ApJ...482..971C, 1998MNRAS.295..743L, 2007ApJ...671..713S}.
Rather, the present simulations are simply meant to illustrate the potential contribution of MSPs to the $\gamma-$ray background based on fiducial population parameters in the literature. 
As will be shown, rather model-independent conclusions can be drawn on the contribution of MSPs to the $\gamma-$ray background and the most accurate estimates are likely based not on constraining the details of the population parameters from radio data, but by calibrating to existing $\gamma-$ray results (\S \ref{comparison with existing gamma rays}).
The parameters we explore are tabulated in Table \ref{msp models table} and the model framework is described below.\\ \\
As a starting point, our models are motivated by the MSP population synthesis study of \cite{1997ApJ...482..971C}.
Another useful reference is the population analysis of the 400 MHz Parkes Southern Pulsar Survey \citep[][]{1998MNRAS.295..743L}.
\cite{2007ApJ...671..713S} also recently performed a MSP population synthesis study.
The important MSP parameters for our purpose are their spatial distribution and their distributions of period, magnetic field, and radio luminosity.\\ \\
Before proceeding, we pause to comment on the binary nature of most MSPs and their subset associated with globular clusters (GCs).
As discussed in \S \ref{radio calibration}, about 80\% of MSPs have a binary companion.
We do not explicitly model binary systems in this work.
Rather, we assume the distributions we prescribe to apply to the MSP population as a whole, including those in binary systems.
The principal effect that is missed in this approach is the reduced sensitivity of surveys to pulsars that are either accelerated or eclipsed in binaries.
As noted in \S \ref{radio calibration}, the unmodelled degradation of the sensitivity to radio MSPs in binary systems leads us to optimistically estimate the sensitivity to MSPs in our simulations and therefore to underestimate the number of MSPs in the Galaxy and their contribution to the $\gamma-$ray sky.
While individual GCs can harbor large concentrations of MSPs, the total number of GC MSPs in the Galaxy is much smaller than the total number of field MSPs. GCs will therefore appear as bright $\gamma-$ray point sources, but will contribute negligibly to the $\gamma-$ray background. We therefore do not model GC MSPs here.\\ \\
FGK06 found the best match to the observed Galactic longitude distribution of ordinary pulsars for a Galactocentric birth radial distribution like peaking at $\approx3$ kpc from the Galactic center \citep[like the one advocated by][]{2004A&A...422..545Y}.
Although this is consistent with the distribution of the massive stellar progenitors of pulsars, MSPs have characteristic spin-down times up to 10 Gyr or more \citep[e.g.,][]{1994ApJ...421L..15C} and so can be generally expected to have gone through many orbits in the Galactic potential.
Their current positions should therefore be largely uncorrelated with their birth locations.
In particular, the annular overdensity and spiral arm structure should be almost completely smeared out.
Thus, for MSPs, we adopt a model in which the surface density projected on the Galactic plane is described by 
\begin{equation}
\rho(r) \propto \exp{(-r^{2}/2\sigma_{r}^{2})};~~~0<r<\infty,
\end{equation}
where $r$ is the distance from the Galactic center.
Based on the birth distribution of ordinary pulsars, a value $\sigma_{r}\sim5$ kpc is expected.\\ \\
The height distribution above and below the Galactic plane is taken to follow a simple exponential distribution,
\begin{equation}
N(z) \propto \exp{(-|z|/\langle |z| \rangle)},
\end{equation}
parameterized by the mean distance from the plane $\langle |z| \rangle$.
For this model, \cite{1997ApJ...482..971C} found $\langle |z| \rangle=0.5$ kpc.
Using dynamical simulations, \cite{1998MNRAS.295..743L} found that the underlying $|z|$ distribution can be quite broad, with a median of 570 pc but a mean of 1.1 kpc, if the velocity distribution is a Maxwellian with a mean 3-D velocity of 130 km s$^{-1}$.
With similar simulations, but with a mean 3-D velocity of 110 km s$^{-1}$ and a different model for the Galactic gravitational potential, \cite{2007ApJ...671..713S} found a $|z|$ scale height of 0.5 kpc, but also with indications of an extended large $|z|$ tail.
We will explore different values from $\langle z \rangle=0.5$ to 2 kpc.\\ \\
\cite{1996MNRAS.283.1383L} studied the underlying spin period distribution of MSPs and highlighted the difference with respect to the observed distribution stemming from the selection effects against detecting the shorter period pulsars.
In their likelihood analysis of these selection effects, \cite{1997ApJ...482..971C} assumed a power-law functional form for the underlying spin period distribution,
\begin{equation}
N(P) \propto  \left\{
\begin{array}{rl}
P^{m_{P,1}-1} & \text{if } P_{\rm min} \leq x < P_{\rm break}\\
P^{m_{P,2}-1} & \text{if }  P \geq P_{\rm break}
\end{array}\right.,
\end{equation}
and found a best fit for $m_{P,1}=m_{P,2}=-1$, with $P_{\rm min}>0.65$ ms at 99\% confidence and $P_{\rm min}>1$ ms at 95\% confidence.
In their analysis a single slope parameter $m_{P}\equiv m_{P,1}=m_{P,2}$ was assumed; we generalize the functional form here to allow for a flattening at short periods.
The parameters $m_{P,1}$ and $m_{P,2}$ correspond to slopes in a log $N$-log $P$ diagram.
The fastest spinning pulsar known at present, J1748$-$2446ad, has a period of 1.4 ms \citep[][]{2006Sci...311.1901H}, constraining $P_{\rm min}$ to be approximately below this value. 
Similarly to the ordinary pulsars, we model the magnetic field distribution of MSPs using a lognormal distribution,
\begin{equation}
N(\log{B}) \propto \exp{[-(\log{B} - \langle \log{B} \rangle)^{2}/2\sigma_{\rm \log{B}}^{2}]},
\end{equation}
with mean logarithm $\langle \log{B} \rangle$ and logarithmic standard deviation $\sigma_{\rm \log{B}}$. 
The period derivative, $\dot{P}$, is then related to the period and magnetic field assuming magnetic dipole braking using the conventional formula $B=3.2\times10^{12}(P\dot{P}/{\rm s})^{1/2}$ G \citep[e.g.,][]{1977QB843.P8M36....}.\\ \\
The radio luminosities of the MSPs, conventionally defined as $L_{\rm r}\equiv S_{\rm r} d^{2}$ (where $S_{\rm r}$ is the flux density at the radio frequency of interest), have been inferred by \cite{1998MNRAS.295..743L} to be well described (after correcting for selection effects) by a log $N$-log $L_{\rm r}$ distribution of the form
\begin{equation}
N(L_{\rm r}) \propto  \left\{
\begin{array}{rl}
L_{\rm r}^{m_{L,1}-1} & \text{if } L_{\rm r,min} \leq L_{\rm r} < L_{\rm r,break}\\
L_{\rm r}^{m_{L,2}-1} & \text{if }  L_{\rm r} \geq L_{\rm r,break}
\end{array} \right.
\end{equation}
with $m_{L}\equiv m_{L,1}=m_{L,2}=-1$, in agreement with ordinary pulsars.
\cite{1997ApJ...482..971C} used the same functional form and found the same slope in their likelihood analysis.
We again have generalized the distribution to allow for a change of slope at $L_{\rm r,break}$.
Contrary to the flattening that is observed for ordinary pulsars below $L_{\rm r}=20$ mJy kpc$^{2}$, \cite{1998MNRAS.295..743L} found a slope $m_{L}=-1$ down to the lowest luminosities sampled $\sim0.3$ mJy kpc$^{2}$ at 400 MHz, suggesting that the number of low-luminosity radio MSPs could exceed that of ordinary pulsars.\\ \\
For consistency with the ordinary pulsar analysis of FGK06, the radio luminosities of the MSPs in our simulations are set at 1.4 GHz.
To convert to luminosities at arbitrary frequencies, each MSP is assigned a spectral index ($S_{\nu}\propto \nu^{-\alpha}$) from a normal distribution with a mean $\langle \alpha \rangle=1.8$ and a standard deviation $\sigma_{\alpha}=0.55$ \citep[][]{1998ApJ...501..270K}.
Like ordinary pulsars, the radio emission from MSPs is assumed to cease when they cross the ``death line'' in the $P-\dot{P}$ diagram (eq. 22 in FGK06).
Because of the short periods of MSPs and since the fastest-spinning ones are assumed to be the strongest $\gamma-$ray emitters, the death line plays a minor role in regulating the cumulative $\gamma-$ray emission from MSPs in our models.
The beaming fractions of MSPs tend to be lower than predicted by fits to the longer-period pulsar population, as in equation (\ref{tauris manchester fb}) \citep[e.g.,][]{1998ApJ...501..270K}, which would imply beaming fractions near unity.
We fiducially take the beaming fraction of MSPs to be $f_{\rm b}^{\rm radio,MSP}=0.5$, though our results can be easily scaled to any other constant value (eq. \ref{beaming scaling}).\\ \\
Of all the pulsar surveys (excluding deep searches toward globular clusters), the MSP detections are dominated by three.
They are the Parkes and Swinburne Multibeam Pulsar Surveys that were also used to constrain the population of ordinary pulsars.
In addition, the Parkes Southern Pulsar Survey which covered all southern hemisphere declinations at 400 MHz \citep{1996MNRAS.279.1235M} took advantage of the steep spectra of pulsars in the radio and was sensitive to low-luminosity pulsars and MSPs.
These surveys detected 20, 21, and 19 MSPs, respectively.
Our MSP population simulations use $2\times10^{6}$ Monte Carlo pulsars each.
Quantities that depend on the total number of pulsars in the Galaxy, such as the $\gamma-$ray background they produce, are calculated by normalizing the number of pulsars to radio detections.
Let $N_{x}^{\rm real}$ and $N_{x}^{\rm sim}$ be the numbers of real and simulated detections for a survey labeled $x$.
We use a normalization weighted by the numbers of detections in the real surveys:
\begin{equation}
\mathscr{N}_{\rm MSP} =
\frac{\sum_{x} N_{x}^{\rm real} (N_{x}^{\rm real}/N_{x}^{\rm sim})}
{\sum_{x} N_{x}^{\rm real}},
\end{equation}
where $x$ runs over the Parkes and Swinburne Multibeam surveys and the Parkes Southern survey.

\subsection{Ordinary vs. Millisecond Pulsars}
\label{ordinary vs msp}
\begin{figure*}
\includegraphics[width=1.0\textwidth]{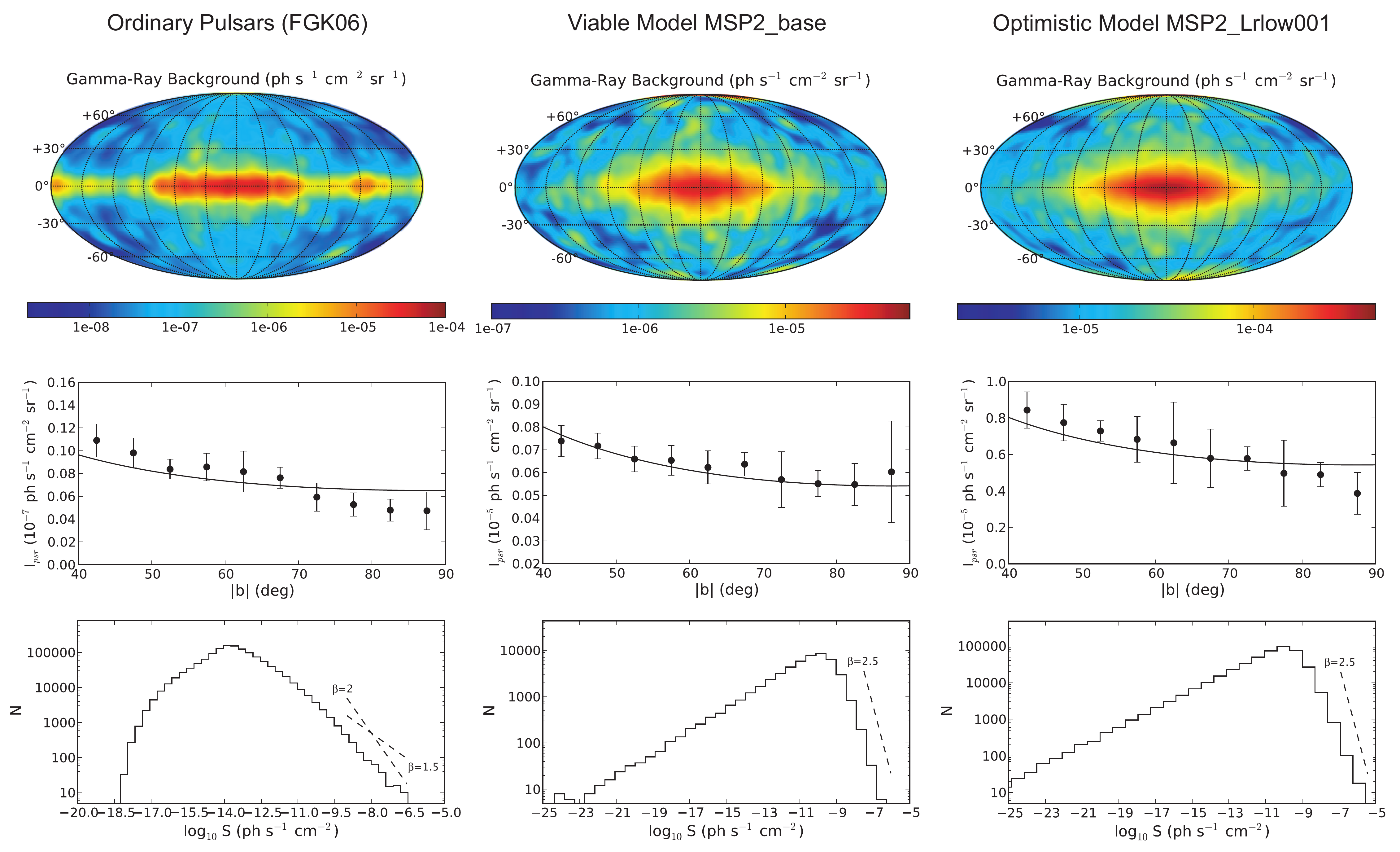}
\caption[]{Detailed results from our $\gamma-$ray calculations.
\emph{Left:} Map of the $\gamma-$ray sky at energies $>$100 MeV, its Galactic latitude profile at $|b|\geq 40^{\circ}$, and the $\log{N}-\log{S}$ flux distribution for our ordinary pulsar model (FGK06).
\emph{Center:} Same, but for a particular MSP model ({\tt MSP2\_base}) selected for its consistency with existing observations (\S \ref{comparison with existing gamma rays}).
\emph{Right:} Same, but for a more optimistic MSP model ({\tt MSP2\_Lrlow001}).
The sky maps were smoothed with a Gaussian of \emph{FWHM} of 4$^{\circ}$, similar to the \emph{EGRET} point spread function.
The solid curves in the middle panels show the best-fit disc-like ($\sim 1/\sin|b|$) latitude profiles.
The dashed lines in the lower panels show different slopes $\beta$ for the high ends of the $\log{N}-\log{S}$ flux distributions, defined such that $dN/dS\propto S^{-\beta}$.
The high-latitude $\gamma-$ray background intensity and \emph{EGRET} and \emph{Fermi} source counts are summarized for all models in Table \ref{msp models table}.
}
\label{maps}
\end{figure*}

In this section, we outline astrophysical considerations that will be helpful in understanding which population, the ordinary pulsars or the MSPs, may be expected to have the greatest contribution to the high-latitude $\gamma-$ray background.
Some of these considerations will also allow us to draw robust, model-independent conclusions with respect to the $\gamma-$ray background contribution of MSPs, for which the population parameters are most uncertain.\\ \\
Let us first focus on the $\gamma-$ray luminosity of individual sources.
Assuming that the $\gamma-$ray luminosities of both populations follow the simple relation in equation (\ref{sqrt Edot prescription}), we can express the ratio of the $\gamma-$ray luminosity of a MSP to that of an ordinary pulsar,
\begin{equation}
\frac{L_{\gamma}^{\rm MSP}}{L_{\gamma}^{\rm ord}}
\sim
\left(
\frac{\dot{P}_{\rm MSP}}{\dot{P}_{\rm ord}}
\right)^{1/2}
\left(
\frac{P_{\rm MSP}}{P_{\rm ord}}
\right)^{-3/2},
\end{equation}
where the subscripts \emph{MSP} and \emph{ord} refer to values characteristic of MSPs and ordinary pulsars, respectively.
In this expression, we have neglected the fact that the $\gamma-$ray luminosity is physically limited to a fraction of the spin-down luminosity of the pulsar.
Taking values $P_{\rm ord}=0.5$ s, $P_{\rm MSP}=3$ ms, $\dot{P}_{\rm ord}=10^{-15}$ s s$^{-1}$, and $\dot{P}_{\rm MSP}=10^{-19}$ s s$^{-1}$, roughly at the center of the observed populations in the $P-\dot{P}$ diagram, we find $L_{\gamma}^{\rm MSP}/L_{\gamma}^{\rm ord}\sim20$.
As a result of their short periods, a ``typical'' MSP may therefore be brighter in the $\gamma-$rays than a ``typical'' ordinary pulsar.
Since $L_{\gamma}\propto \sqrt{\dot{E}}$ in this model, the conclusion remains qualitatively true even if the $\gamma-$ray luminosity is limited at a fraction of $\dot{E}$.
\\ \\
Note that the sample of observed ordinary pulsars may be biased toward short-period pulsars if young objects tend to be brighter in the radio (e.g., FGK06).
On the other hand, the most rapidly rotating MSPs are strongly disfavored in existing surveys.
It is therefore likely that the known ordinary pulsars are biased toward the $\gamma-$ray bright end of their population, whereas the known MSPs may be biased toward their $\gamma-$ray faint end.
As a result, the truly typical MSP may be brighter in the $\gamma-$rays by a substantially larger factor than the truly typical ordinary pulsar than estimated above.\\ \\
The above argument is perfectly consistent with the few \emph{observed} ordinary pulsars to appear more luminous in the $\gamma-$rays than the few \emph{observed} MSPs, as suggested by Figure \ref{Lgamma vs Edot}.
In fact, the brightest members of these respective populations are not representative of their typical members.
Any meaningful comparison must take into account the very different luminosity distributions of these populations.
We have explicitly verified using our population models that the $\gamma-$ray luminosities of the ordinary pulsars predicted to have been
\emph{detected} in the early \emph{Fermi} data are generally higher than the same quantity for the \emph{detected} MSPs, in agreement with the data shown in Figure \ref{Lgamma vs Edot}.
At the same time, the \emph{average} member of the underlying MSP population is predicted to be more luminous than the \emph{average} member of the underlying ordinary pulsar population, and MSPs are predicted to provide a higher contribution to the $\gamma-$ray background.
This finding strongly cautions against drawing conclusions from the observed ``tip of the iceberg'' of the $\gamma-$ray pulsars.\\ \\
How many Galactic MSPs are there compared to ordinary pulsars?\\ \\
If MSPs are recycled pulsars, then averaged over a long time scale the birthrate of MSPs must necessarily be less than the birthrate of ordinary pulsars, which may viewed as the rate of ``pulsar creation.''
Based on this consideration only, the MSP birthrate can be expected to be substantially lower than that of ordinary pulsars since only pulsars in binary systems with specific properties \citep[like low-mass X-ray binaries, LMXBs; e.g.,][]{1991PhR...203....1B} can recycle them.
Owing to their magnetic fields suppressed by a factor $\sim10^{3}-10^{5}$, MSPs however have smaller period derivatives $\dot{P}\propto B^{2}/P$, and hence longer spin-down time scales $\tau_{c}\sim P/\dot{P} \propto (P/B)^{2}$.
The longer lifetimes of MSPs can therefore compensate for their lower birthrate so that the number of active Galactic MSPs can be comparable or even higher than that of ordinary pulsars.
\cite{1998MNRAS.295..743L} in fact estimate that the number of Galactic MSPs beamed toward us in the radio is about equal to the number of ordinary pulsars, also beamed toward us, for luminosities above 1 mJy kpc$^{2}$ at 400 MHz.
While the luminosity function of ordinary pulsars appears to flatten at luminosities less than about 1 mJy kpc$^{2}$, these authors found no evidence for a flattening of the MSP luminosity function down to the lowest  luminosities sampled of $\sim0.3$ mJy kpc$^{2}$, suggesting observationally that the underlying population of MSPs could in fact exceed that of ordinary pulsars.\\ \\
As we are principally interested in the high-latitude $\gamma-$ray background from pulsars that may be confused with the extragalactic component, another key aspect to consider is the Galactic latitude distribution of the $\gamma-$ray emission from ordinary pulsars versus MSPs.
Whereas young, energetic ordinary pulsars are born in and are concentrated close to the Galactic plane, the ages of MSPs generally exceed the oscillation time across the plane of the Galaxy ($\sim100$ Myr) by a large factor.
Therefore, the height of a MSP above the Galactic plane should be uncorrelated with its rotational parameters.
As a consequence, \emph{while the $\gamma-$ray bright ordinary pulsars are concentrated near the plane, the $\gamma-$ray bright MSPs should have a scale height that is more representative of the parent population as a whole, and thus be more prevalent at high latitudes}.
This observation explains why the $\gamma-$ray flux distribution of MSPs is predicted to be close to Euclidean at the high end, in contrast to ordinary $\gamma-$ray pulsars which have a more planar distribution (Fig. \ref{maps}).
The simple fact that the shape of the flux distribution is determined predominantly by geometrical considerations (see Appendix \ref{slope of the flux distribution}) allows us to place largely model-independent constraints on the contribution of MSPs to the $\gamma-$ray background (\S \ref{comparison with existing gamma rays}).\\ \\

\nopagebreak

\subsection{Model Results}
\label{model results}
Using the Galactic pulsar population models and the $\gamma-$ray luminosity prescription described in \S \ref{population synthesis}$-$\ref{msps}, we are now ready to calculate the $\gamma-$ray background they produce.
For each model, we calculate the average cumulative $\gamma-$ray photon intensity ($>$100 MeV) from pulsars with absolute Galactic latitude $|b|$ greater than a minimum value $b_{\rm min}$:  
\begin{equation}
\label{background intensity eq}
I_{\rm psr}(|b|\geq b_{\rm min}) =
\frac{1}{\Omega{(|b|\geq b_{\rm min}})} 
\sum_{|b|\geq b_{\rm min}} 
\frac{L_{\gamma}^{\rm ph}}{4 \pi d^{2}},
\end{equation}
where $\Omega{(|b|\geq b_{\rm min}})$ is the solid angle subtended by latitudes $|b|\geq b_{\rm min}$ and the sum is over all the pulsars within that solid angle.
Since we are interested in the high-latitude background that could be confused with the extragalactic component, we set $b_{\rm min}=40^{\circ}$.
Pulsar fluxes in equation (\ref{background intensity eq}) assume a $\gamma-$ray beaming fraction of unity, so that they spread over a solid angle $4\pi$, consistent with the $\gamma-$ray luminosity calibration of \S \ref{observed gamma ray luminosities}.
Note, however, that this calibration was based on observed fluxes and that the beaming fraction factors cancel out exactly in the flux calculation of each pulsar.
As discussed below, the $\gamma-$ray background however scales with the fraction of pulsars that have $\gamma-$ray beams encompassing our line of sight to them.
In what follows, we will simply write $I_{\rm psr}$ for $I_{\rm psr}(|b|\geq 40^{\circ})$. \\ \\
The results of our population syntheses are given in Table \ref{msp models table}.
For the ordinary pulsars, the model parameters are fixed to the ones constrained using radio observations by FGK06.
For this model, ordinary pulsars are predicted to produce a small fraction, $\sim10^{-3}$, of the high-latitude $\gamma-$ray background above 100 MeV.
A range of parameters are explored for the MSP population.
In \S \ref{comparison with existing gamma rays}, we will compare the different models with existing $\gamma-$ray observations, including the $\gamma-$ray background and early \emph{Fermi} sources.
In the remainder of this section, we first explore how $I_{\rm psr}$ depends on model parameters for the MSPs.\\ \\
{\bf Radial distribution.} In our models, the Sun is at a distance $R_{\odot}=8.5$ kpc from the Galactic center \citep[][]{1997MNRAS.291..683F, 1998ApJ...494L.219P, 2009arXiv0902.3913R}.
We have explored values $\sigma_{r}=3,~5,~{\rm and}~7$ kpc (e.g., 
models {\tt MSP1\_base, MSP1\_sigr3, MSP1\_sigr7}) and for these $I_{\rm psr}$ increases with $\sigma_{r}$, presumably because the bulk of the MSPs is located closer to us.\\ \\
{\bf Scale height.} For the scale height of MSPs, we have tried values of $\langle |z| \rangle=0.5,~1,~1.5,~{\rm and}~2$ kpc (e.g., 
models {\tt MSP1\_base, MSP1\_z05, MSP1\_z15, MSP1\_z20}) and find that $I_{\rm psr}$ increases with $\langle |z| \rangle$.
This is expected, as more objects lie at high latitudes for larger $\langle |z| \rangle$.\\ \\
{\bf Period distribution.} The $\gamma-$ray background is particularly sensitive to the period distribution of MSPs, which is not surprising given the adopted scaling of the $\gamma-$ray luminosity of each pulsar with its period in equation (\ref{sqrt Edot prescription}).
For a logarithmic slope $m_{P,1}=m_{P,2}=-1$ of the underlying period distribution, as found by \cite{1997ApJ...482..971C}, $I_{\rm psr}$ increases strongly with decreasing minimum period $P_{\rm low}$ as the abundance of rapidly-rotating, energetic MSPs increases (e.g., 
models {\tt MSP1\_base, MSP1\_Plow065, MSP1\_Plow125, MSP1\_Plow15}).
We have also explored cases in which the period distribution flattens to $m_{P,1}=0$ at periods below a break period $P_{\rm break}$.
For these cases, $I_{\rm psr}$ increases with decreasing $P_{\rm break}$ at fixed $P_{\rm low}$ (e.g., 
models {\tt MSP1\_Pbreak15, MSP1\_Pbreak3, MSP1\_Pbreak5, MSP1\_Pbreak10}), again because the abundance of short-period pulsars in enhanced.\\ \\
{\bf Magnetic field distribution.} The magnetic field distribution is similarly important for the $\gamma-$ray background as the $\gamma-$ray luminosity is also assumed to be a function of the period derivative.
In the magnetic dipole spin-down model, the period and period derivative are directly related to the magnetic field by $B\propto \sqrt{P \dot{P}}$ \citep[][]{1969ApJ...157.1395O, 1969ApJ...157..869G, 2006ApJ...648L..51S}.
A $\gamma-$ray luminosity law $L_{\gamma}\propto \dot{P}^{1/2} P^{-3/2}$ can therefore be expressed as $L_{\gamma}\propto B P^{-2}$.
As expected, $I_{\rm psr}$ then increases with $\langle \log{B} \rangle$ (e.g.,
models {\tt MSP1\_base, MSP1\_logB8, MSP1\_logB825, MSP1\_logB875}).
At fixed $\langle \log{B} \rangle$, it also increases with increasing $\sigma_{\log{B}}$ as the tail of high $B$ pulsars becomes more extended (e.g., 
models {\tt MSP1\_base, MSP1\_sigB03, MSP1\_sigB04, MSP1\_sigB05}).\\ \\
{\bf Radio luminosity distribution.} The luminosity function of MSPs in the radio is also important for the cumulative $\gamma-$ray emission predicted by the population synthesis models calibrated based on radio detections (\S \ref{radio calibration}).
In models for which the radio luminosity function of MSPs extends to very low luminosities, the Galaxy can contain a large number of MSPs that are undetected in the radio, yet contribute to the $\gamma-$ray background.
For a MSP radio luminosity function with logarithmic slope $m_{L,1}=-1$ \citep[e.g.,][]{1998MNRAS.295..743L}, $I_{\rm psr}$ thus increases with decreasing low-luminosity cutoff $L_{\rm r,low}$ (e.g., 
models {\tt MSP1\_base, MSP1\_Lrlow001, MSP1\_Lrlow10}).
Since the $\gamma-$ray background could in principle be arbitrarily boosted by MSPs which have very low radio luminosities, the background can be used to limit the abundance of the radio-faint sources.
This is however not an issue for our more physical ordinary pulsar model, which was constrained in details in FGK06.
Moreover, existing $\gamma-$ray observations, combined with the robust prediction for the shape of the MSP $\gamma-$ray flux distribution, constrain their contribution to the $\gamma-$ray background relatively independently of the model details (see \S \ref{comparison with existing gamma rays} and Strong 2007\nocite{2007Ap&SS.309...35S} for a related approach concerning the Galactic diffuse emission).\\ \\
{\bf Beaming fractions.} Our calculations assume a beaming fraction $f_{\rm b}^{\rm radio,MSP}=0.5$ for MSPs in the radio and $f_{\rm b}^{\gamma}=1$ in the $\gamma-$rays.
While the $\gamma-$ray background scales with the fraction of pulsars emitting $\gamma-$rays in our direction, it scales inversely to the fraction whose radio beams intersect our sightline as a result of the calibration to the number of detections in radio surveys:
\begin{equation}
\label{beaming scaling}
I_{\rm psr} = 
I_{\rm psr}^{\rm fid}
\left( \frac{f_{\rm b}^{\gamma}}{1} \right)
\left( \frac{f_{\rm b}^{\rm radio,MSP}}{0.5} \right)^{-1},
\end{equation}
where $I_{\rm psr}^{\rm fid}$ is the intensity obtained using the fiducial beaming fractions and reported in Table \ref{msp models table}.\\ \\
Figure \ref{maps} shows detailed results from our $\gamma-$ray calculations.
The left column shows a map of the $\gamma-$ray sky, its Galactic latitude profile at $|b|\geq 40^{\circ}$, and the $\log{N}-\log{S}$ flux distribution for our ordinary pulsar model.
The solid curve in the middle panel shows the best-fit disc-like ($\sim 1/\sin|b|$) latitude profile.
The central column shows the same quantities, but for a particular MSP model ({\tt MSP2\_base}) selected for its consistency with existing observations, to be discussed in the next section.
The right column also shows the same quantities but for a more optimistic MSP model ({\tt MSP2\_Lrlow001}).
For the sky maps, sources with flux $S>10^{-7}$ ph s$^{-1}$ cm$^{-2}$ (the approximate point-source sensitivity of \emph{EGRET}; see \S \ref{egret sources and fermi predictions}) were omitted to avoid bright spots corresponding to resolved sources.
The high-latitude $\gamma-$ray background intensity is given in Table \ref{msp models table} for all models explored.

\section{Comparison with Existing Gamma-Ray Observations}
\label{comparison with existing gamma rays}
Next, we explore the constraints that existing $\gamma-$ray observations put on the contribution of pulsars to the $\gamma-$ray background.

\subsection{Existing Detections}
\label{egret sources and fermi predictions}
An important test of our $\gamma-$ray pulsar population models is provided by the number of detections by \emph{EGRET} and early \emph{Fermi} observations.
Specifically, the number of high $\gamma-$ray flux pulsars predicted by a given model must be consistent with existing \emph{EGRET} and \emph{Fermi} results for the model to be viable.
The strongest constraints are provided by the first three months \emph{Fermi} catalog \citep[][]{2009ApJS..183...46A}, in which a larger fraction of the sources are identified than in the \emph{EGRET} data, and so we focus on these.\\ \\
Unfortunately, the comparison is complicated by the complex selection effects in detecting $\gamma-$ray pulsars.
For example, three different approaches to detecting pulsars are adopted by \emph{Fermi} \citep[][]{2009arXiv0901.3405S}.
First, $\gamma-$ray pulsations are searched for toward pulsars that are already known, usually in the radio.
Because $\gamma-$ray sources are inherently dim, it is easier to detect pulsations when their period is known from measurements at other wavelengths.
Second, blind searches for $\gamma-$ray pulsars are carried out using a time-differencing technique toward unidentified \emph{EGRET} and \emph{Fermi} sources.
These searches are important since many pulsars appear to be radio-quiet (e.g., the Geminga and CTA 1 pulsars) and may only be detectable in the $\gamma-$rays.
They are however generally less sensitive when the period is not known a priori and currently limited to $P>16$ ms and $\dot{P}/P<1.25\times10^{-11}$ s$^{-1}$ \citep[][]{2009arXiv0901.3405S}.
Third, many of the blind searches actually target well-studied astrophysical sources (including \emph{EGRET} unidentified sources, supernova remnants, and TeV sources discovered by ground-based Cerenkov telescopes) which, to various degrees of confidence, are suspected to contain pulsars.
The detection threshold is also a function of position on the sky, with the diffuse background being stronger near the Galactic plane \citep[e.g.,][]{2009ApJS..183...46A}.\\ \\
To avoid the large uncertainties from the complicated selection function, we therefore base our discussion on the flux complete subset of the analysis of the $|b|\geq10^{\circ}$ sources reported by \cite{2009ApJ...700..597A}.
This corresponds to a flux threshold of $S>1.25\times10^{-7}$ ph s$^{-1}$ cm$^{-2}$ at energies $>$100 MeV.
We denote the number of simulated sources with flux above this threshold and with latitude $|b|\geq 10^{\circ}$ by $N_{\rm comp}(|b|\geq10^{\circ})$.
Among the sources with flux above this threshold, two are new pulsars discovered by \emph{Fermi} (LAT PSR J1836+5925 and LAT PSR J0007+7303) and two more are unidentified.
While 40 are associated at high-confidence with AGNs, the high-confidence associations are only defined as having probability $>$90\% of being genuine, so that a few could be spurious.
In addition, PSR J0357.5+3205 is a previously-known MSP with $\gamma-$ray flux just below the completeness threshold, but consistent with a value above the threshold within the uncertainty.
Because the $\gamma-$ray beams of pulsars appear generally wider than their radio beams and because many pulsars have not yet been detected at other wavelengths, we can expect at least as many $\gamma-$ray pulsars to be unidentified as sources with known counterparts.
It is thus likely that the sources in the flux limited sample that are not associated with a known AGN are pulsars.
Conservatively and accounting for the fact that some of the potential sources may not be MSPs, we fiducially  take the allowed simulated $N_{\rm comp}(|b|\geq10^{\circ})$ to be 4, with a significant statistical uncertainty.
At the $2\sigma$ level, 8 sources are allowed.
The predicted $N_{\rm comp}(|b|\geq10^{\circ})$ for each model is given in Table \ref{msp models table}\\ \\

\begin{table*}
\centering
\caption{Extragalactic $\gamma$-ray ($>$100 MeV) Background Estimates\label{bkg measurements}}
\begin{tabular}{cccc}
\hline\hline
Reference                                 & $I_{X}$                                     & Spectrum & Instrument  \\
                                          & 10$^{-5}$ ph s$^{-1}$ cm$^{-2}$ sr$^{-1}$   &          &             \\
\hline
\cite{tf82} & $1.3\pm0.5$                             & $p=2.35^{+0.4}_{-0.3}$                                 & \emph{SAS-2} \\
\cite{1994JPhG...20.1089O}    & $1.10\pm0.05$                          & $p=2.11\pm0.05$\tablenotemark{a}          & \emph{EGRET} \\
\cite{1998ApJ...494..523S}    & $1.45\pm0.05$                          & $p=2.10\pm0.03$\tablenotemark{a}          & \emph{EGRET} \\
\cite{2004JCAP...04..006K}    & $<0.5$ (99\% conf.)\tablenotemark{b}   & $-$                      & \emph{EGRET} \\
\cite{2004ApJ...613..956S}    & $1.11\pm0.01$                          & positive curvature\tablenotemark{c}       & \emph{EGRET} \\
\hline

\end{tabular}

\tablenotetext{a}{Power-law photon index.}
\tablenotetext{b}{These authors found that the $\gamma$-ray background Galactic latitude profile is disc-like ($\sim 1/\sin{|b|}$) for $40^{\circ} \lesssim |b| \lesssim 70^{\circ}$ and even steeper for $|b| \gtrsim 70^{\circ}$, with an average intensity at the Galactic poles ($|b|>86^{\circ}$) $I_{\rm pole}=1.20\pm0.08\times10^{-5}$ ph s$^{-1}$ cm$^{-2}$ sr$^{-1}$ before Galactic foreground subtraction.}
\tablenotetext{c}{Not well described by a power law. The positive curvature may be affected by systematic uncertainties.}

\end{table*}

\subsection{Background Intensity and Latitude Profile}
\label{background intensity}
The extragalactic $\gamma-$ray background is usually measured by subtracting from the total intensity a Galactic foreground model accounting for $\gamma-$rays produced by cosmic-ray nucleons interacting with nucleons in the interstellar gas, bremsstrahlung by cosmic-ray electrons, and inverse Compton interaction of cosmic-ray electrons with ambient low-energy interstellar photons \citep[e.g.,][]{1998ApJ...494..523S, 2004ApJ...613..956S}.
A potential pulsar contribution would therefore be lumped into the extragalactic component $I_{X}$.
Existing extragalactic background estimates can therefore be used to constrain the Galactic pulsar population and its $\gamma-$ray emission from the requirement that their integrated $\gamma-$ray intensity must not exceed these.
Table \ref{bkg measurements} lists existing estimates of the extragalactic $\gamma-$ray background above 100 MeV, along with the corresponding spectral shape in the cases where it was also measured.\\ \\
The original analysis of the final $\emph{EGRET}$ data by \cite{1998ApJ...494..523S} found $I_{X}=(1.45\pm0.05)\times10^{-5}$ ph s$^{-1}$ cm$^{-2}$ sr$^{-1}$, while a more recent analysis by \cite{2004ApJ...613..956S} using an updated foreground model found a lower value $I_{X}=(1.11\pm0.01)\times10^{-5}$ ph s$^{-1}$ cm$^{-2}$ sr$^{-1}$.
At present about 7\% of the extragalactic background has been resolved into blazars by early \emph{Fermi} observations \citep[][]{2009ApJ...700..597A}, though this fraction is likely to increase as deeper integrations are obtained.
An upper limit to the pulsar contribution is thus $I_{\rm psr}\lesssim1\times10^{-5}$ ph s$^{-1}$ cm$^{-2}$ sr$^{-1}$.\\ \\
In a more model-independent analysis based on correlations with Galactic tracers and on the Galactic latitude profile of the $\gamma-$ray background, \cite{2004JCAP...04..006K} find $I_{X}<0.5\times10^{-5}$ ph s$^{-1}$ cm$^{-2}$ sr$^{-1}$ at the 99\% confidence level, with evidence for an even lower intensity.
The discrepancy between the \cite{2004JCAP...04..006K} analysis and the measurements of \cite{1998ApJ...494..523S} and \cite{2004ApJ...613..956S} could be understood if pulsars were picked up as a Galactic component in the \cite{2004JCAP...04..006K} analysis.  
The $\gamma-$ray Galactic latitude profiles predicted for our pulsars models (Figure \ref{maps}), with their nearly disc-like behavior $\sim 1/\sin|b|$, are in fact tantalizingly similar to the one measured at $|b|\geq40^{\circ}$ \citep[see Figure 3 of][]{2004JCAP...04..006K}.

\subsection{Background Spectrum}
\label{background spectrum}
The sources that dominate the isotropic component of the $\gamma-$ray background must have spectra that are consistent with the latter.
While there remains significant variance between the different estimates (Table \ref{bkg measurements}), these indicate a spectrum with approximately equal power per energy decade (a power-law photon index $p\approx2$), or slightly steeper and possibly with positive curvature, between 100 MeV and 100 GeV.
In fact, the similar spectra of blazars were used as an argument in favor of attributing most of the isotropic component to these sources \citep[][]{1998ApJ...494..523S, 2004ApJ...613..956S}.\\ \\
Are the $\gamma-$ray spectra of pulsars also compatible with them making a significant contribution to the high-latitude $\gamma-$ray background?
The relatively flat spectra of pulsars, up to a certain energy, are indeed in agreement with the \emph{EGRET} $\gamma-$ray background spectrum measurements.
However, of the few pulsars with $\gamma-$ray spectra, many exhibit high-energy cutoffs around a few GeV.
For instance, the spectrum of the Vela pulsar is exponentially suppressed at around 2.9 GeV  \citep[][]{2009ApJ...696.1084A}.
The first MSP with a spectrum measured by \emph{Fermi}, PSR J0030+045, also exhibits a cut off at around 1.7 GeV \citep{2009ApJ...699.1171A}.
Such a pronounced cutoff is not observed in the $\gamma-$ray background spectrum measured by \cite{1998ApJ...494..523S} at these energies and would place a strong constraint on the contribution of pulsars if it were typical of the pulsars dominating the background component.
\emph{Fermi} measurements of the $\gamma-$ray spectra of more MSPs will therefore be extremely valuable in quantifying their contribution to the $\gamma-$ray background.\\ \\
Note, however, that in their analysis of the same \emph{EGRET} data with an improved Galactic cosmic-ray foreground model \cite{2004ApJ...613..956S} measure a residual $\gamma-$ray background spectrum that is inconsistent with both a simple power law and with the original \cite{1998ApJ...494..523S} analysis.
The \cite{2004ApJ...613..956S} measurement is in fact indicative of an upturn in the spectrum around a few GeV.
It is thus possible that pulsars contribute a significant fraction of the background up to energies of a few GeV even if their spectra are suppressed at these energies, if other sources provide the photons observed at higher energies.
It will be necessary to also obtain a firmer measurement of the background spectrum before stronger constraints can be put on the contribution of pulsars to this background from measurements of their individual spectra.\\ \\
A preliminary analysis of the \emph{Fermi} data (see the presentation by M. Ackermann at http://www-conf.slac.stanford.edu/tevpa09/) indicates an isotropic $\gamma-$ray background spectrum that is steeper than the original \emph{EGRET} analysis by \cite{1998ApJ...494..523S} with $p\approx 2.45$ between 200 MeV and 50 GeV, and of lower intensity, in better agreement with the \cite{2004ApJ...613..956S} normalization. 
As the error bars converge, the presence or absence of a feature in this spectrum around few GeV will strongly constrain the contribution of MSPs if their spectra are generically suppressed at these energies. 
The preliminary \emph{Fermi} error bars at a few GeV, where the observed pulsar spectra peak relative to the background spectrum, are large enough to comfortably accommodate a $5-15$\% contribution from MSPs, as allowed by other constraints (\S \ref{which models pass tests}).

\subsection{Which Models Pass the Tests?}
\label{which models pass tests}
Among the models explored (Table \ref{msp models table}), which ones are compatible with the existing data?
The basic requirements are: 1) $N_{\rm comp}(|b|\geq10^{\circ}) \lesssim 4$ and 2) $I_{\rm psr}\lesssim1\times10^{-5}$ ph s$^{-1}$ cm$^{-2}$ sr$^{-1}$.\\ \\
With a fractional high-latitude $\gamma-$ray background contribution of $\sim10^{-3}$, our model for ordinary pulsars appears quite consistent with the data.
Because the $\gamma-$ray bright ordinary pulsars are concentrated near the Galactic plane, the $|b|>10^{\circ}$ data provide weak constraints.
We have checked that the model predicts only a total of 17 sources on the sky above a flux threshold of $1\times10^{-7}$ ph s$^{-1}$ cm$^{-2}$, roughly the average \emph{EGRET} point source sensitivity\footnote{Near the Galactic plane, the point source detection threshold is somewhat higher owing to the background, so this estimate is optimistic.} \citep[][]{1999APh....11..277G}.
This is in good agreement with the six pulsars detected by \emph{EGRET} and the 170 unidentified sources in the third catalog \citep[][]{1999ApJS..123...79H}.
Most of our MSP models can however already be excluded on the basis that they overproduce either $N_{\rm comp}(|b|\geq10^{\circ})$ or the $\gamma-$ray background.
There are however viable and interesting models.
Model {\tt MSP2\_base} (with $P_{\rm low}=1.5$ ms and $\langle \log{B} \rangle=8.0$), whose detailed results are plotted in Figure \ref{maps}, predicts $N_{\rm comp}(|b|\geq10^{\circ})=4$ and a high-latitude $\gamma-$ray background from MSPs $I_{\rm psr}\approx0.08\times10^{-5}$ ph s$^{-1}$ cm$^{-2}$ sr$^{-1}$.
This corresponds to a sizable fraction of the extragalactic value: slightly less than $\sim10$\% if the measurements obtained by subtracting a cosmic ray model are correct, but $\sim 20$\% or more if the more model-independent analysis of \cite{2004JCAP...04..006K} is more accurate (Table \ref{bkg measurements}).
It is interesting that the MSPs that are the most luminous in the $\gamma-$rays are likely to be those with the highest $\dot{E}$ and therefore the most likely to ablate their companions and be eclipsed by the vaporized material.
Such pulsars, which would be difficult to identify in the radio or in blind $\gamma-$ray searches, could remain $\gamma-$ray sources either by transmitting most of their intrinsic $\gamma-$ray luminosity or by secondary $\gamma-$rays produced in the interaction with the obscuring material \citep[][]{1991ApJ...379L..69T}.
The more optimistic MSP model {\tt MSP2\_Lrlow001}, whose detailed results are shown in the right column of Figure \ref{maps}, overpredicts the number early \emph{Fermi} detections, but illustrates how MSPs could potentially contribute an even large fraction of the background under reasonable assumptions.\\ \\
The fact that many of our MSP models, with quite reasonable population parameters based on the literature, overproduce the high-latitude $\gamma-$ray background is particularly notable, as it implies that the background puts interesting constraints on the global Galactic MSP population.
We currently only know of $\sim 100$ MSPs in the Galactic field, although there could be $\sim 10^{5}$ or more in total \citep[][]{1998MNRAS.295..743L}, many of which would be undetectable in the radio because of beaming.  
Although this may suggest that our MSP $\gamma-$ray background predictions are wildly uncertain, this is not the case.
In general, the number of resolved point sources above a given flux threshold is not independent of the integrated flux from the parent population.
In Appendix \ref{relation between resolved sources and the background}, we derive this relation and  analytically estimate the contribution of MSPs to the high-latitude $\gamma-$ray background from the number of resolved point sources in the early complete $\emph{Fermi}$ sample.
We find $I_{\rm psr} \sim 0.125 \times 10^{-5}{\rm~ph~s^{-1}~cm^{-2}~sr^{-1}}$ for $N_{\rm comp}(|b|\geq10^{\circ})=4$, in good agreement with our numerical calculations for models with consistent numbers of resolved sources.
This estimate relies on the slope $\beta$ of the $\log{N}-\log{S}$ distribution (defined such that $dN/dS \propto S^{-\beta}$), which is empirically found to be nearly Euclidean ($\beta=2.5$) at the high-flux end (Figure \ref{maps}).
In Appendix \ref{slope of the flux distribution}, we show that this slope should be robust to the details of the population models,
as we have empirically verified to be the case for all of our MSP models.
It is also worth noting that \cite{2007ApJ...671..713S}, who performed more extensive calibrations to radio data for MSPs and treated beaming geometries in great detail, predict a nearly identical shape for the $\gamma-$ray flux distribution (see their Fig. 10).
Note again that both the small-number statistical uncertainty on $N_{\rm comp}(|b|\geq10^{\circ})$ and the possibility of a higher value owing to spurious associations of some sources with AGNs might allow an even more significant contribution of MSPs to the $\gamma-$ray background.

\section{Implications for Pulsar Physics}
\label{physics implications}
The previous section was chiefly concerned with the constraints that can be put on the pulsar population as whole, particularly MSPs, using the $\gamma-$ray background.
These also have implications for the physical processes that govern pulsars.

\subsection{Maximum Spin Frequency}
If $\gamma-$ray emission from pulsars, including MSPs, is in fact a function of the period and period derivative of the pulsar of the type in equation (\ref{sqrt Edot prescription}), then the measured $\gamma-$ray background puts a lower limit on the maximum spin frequency (or minimum spin-period $P_{\rm low}$), as argued \S \ref{comparison with existing gamma rays}.
This has interesting consequences arising from the physics that limit this spin frequency: the break-up spin rate, accretion spin-magnetic equilibrium, and gravitational wave emission.
We discuss these below.\\ \\
A strict upper limit on the spin frequency of neutron stars is the angular velocity at which centrifugal acceleration would disrupt the star, usually called the ``break-up limit.''
This break-up frequency depends on the equation of state of the ultradense matter at the cores of neutron stars.
A firm upper limit of $\approx3$ kHz is set by the requirement that the sound speed within the star cannot exceed the speed of light.
More realistic equations of state predict lower maximum spin rates in the 1500$-$2000 Hz range \citep[][]{1994ApJ...424..823C, 1999A&A...344..151H, 2001ApJ...550..426L, 2008AIPC.1068...67C}.
In order to accommodate a radius of 8$-$12 km for a 1.4 M$_{\odot}$ neutron star \citep[e.g.,][]{1999ApJ...512..288T, 2006Natur.441.1115O, 2008arXiv0811.3979G, 2009ApJ...693.1775O}, a break-up limit greater than about 1500 Hz is needed.\\ \\
It took 25 years from the discovery of the first MSP, B1937+21 with a spin period of 1.6 ms, before a more-rapidly rotating pulsar, J1748$-$2446ad  with a period of 1.4 ms, was found \citep[][]{2006Sci...311.1901H}.
While \cite{2006Sci...311.1901H} argue that the difficulty of detecting such a pulsar suggests that more with comparable or shorter periods may exist in the Galaxy, observations of nuclear-powered millisecond X-ray pulsars, which during outbursts are bright enough to be detected throughout the Galaxy and are not affected by the same selection effects, however indicate that 1.4 ms is close the minimum spin period \citep[e.g.,][]{2008AIPC.1068...67C}.
The fact that this putative minimum spin period is a factor of two or more above the break-up limit for favored equations of state is interesting as it may be evidence that a different mechanism is responsible for the dearth of pulsars spinning at higher rates.\\ \\
The most conservative explanation for a maximum spin frequency above the break-up limit is rooted in the origin of MSPs in LMXBs.
In this recycling scenario, the binary companion to the neutron star fills its Roche lobe and transfer of orbital angular momentum to spin angular momentum ensues from accretion of material onto the compact object \citep[e.g.,][]{1991PhR...203....1B}.
Because the neutron star is strongly magnetized, the Alfven radius $R_{\rm A}$, at which the kinetic energy density of the accreting material becomes comparable to the local magnetic energy density, is well outside of the neutron star surface.
As a result, the accretion flow exerts viscous torques at $R_{\rm A}$ until an equilibrium spin period, approximated by the angular Keplerian velocity at $R_{\rm A}$, is reached: 
\begin{equation}
P_{\rm eq} \approx 1.9~{\rm ms}
\left( \frac{B}{10^{9}~{\rm G}} \right)^{6/7}
\left( \frac{\dot{M}}{\dot{M}_{\rm Ed}} \right)^{-3/7}
\left( \frac{R_{\star}}{10~{\rm km}} \right)^{16/7},
\end{equation}
where $\dot{M}/\dot{M}_{\rm Ed}$ is the mass accretion rate as a fraction of the Eddington limit and $R_{\star}$ is the radius of the neutron star \citep[e.g.,][]{1992xbfb.work..487G}.
A short-period cutoff that is the result of magnetic spin equilibrium would therefore constrain the joint distribution of $B$, $\dot{M}$, and $R$.
A sharp period cutoff seems unlikely in this scenario as, for example, there is no obvious reason for $B$ to be strictly limited to a particular range; rather, the $P-\dot{P}$ diagram is suggestive of a broad $B$ distribution.\\ \\
Another interesting possibility is that accretion spin-up torques may be mitigated by angular momentum losses owing to gravitational radiation.
These could occur via a number of proposed mechanisms, including $r-$modes or a quadrupole moment induced either by accretion or a toroidal magnetic field \citep[e.g.,][]{1984ApJ...278..345W, 1999ApJ...516..307A, 2000MNRAS.319..902U, 2002PhRvD..66h4025C, 2005ApJ...623.1044M}.
An attractive aspect of gravitational wave braking is that the spin-down torque scales with the fifth power of the spin frequency, $\propto \omega^{5}$, and would thus more naturally produce a sharp suppression at short spin periods.\\ \\
If \emph{Fermi} observations can provide a reliable $\gamma-$ray emission law for MSPs, then population synthesis models like the ones we have explored could be used to better constrain the short-end of the period distribution using the $\gamma-$background and thus help determine the physics that limits pulsar spin rates.

\subsection{Gamma-Ray Emission Mechanisms}
\label{gamma ray emission mechanisms}
The most obvious question, of course, is how can the $\gamma-$ray background inform us on the pulsar $\gamma-$ray emission mechanisms?
Two main classes of $\gamma-$ray emission mechanisms, reviewed by \cite{2007Ap&SS.309..221H}, have been proposed.
Polar cap models assume that particle acceleration takes place near the neutron star magnetic poles and that $\gamma-$ray emission results from cascades initiated by magnetic pair production in the strong magnetic field \citep[e.g.][]{1982ApJ...252..337D, 1996ApJ...458..278D}.
Outer gap models instead assume that the particle acceleration takes place in vacuum gaps that form in the outer magnetosphere \citep[e.g.,][]{1986ApJ...300..500C, 1996ApJ...470..469R}. 
The slot gap model is a variant of the polar cap model in which acceleration occurs in a `slot gap', a narrow region bordering the last open field line in which the electric field is unscreened.
The high-energy emission beam of the slot gap cascade is a hollow cone with a much larger opening angle than that of the polar cap cascade emission.\\ \\
$\gamma-$ray observations provide a few handles on the pulsar emission mechanisms. 
While our empirical approach to modelling $\gamma-$ray emission does not allow us to make quantitative statements, we briefly outline discriminating statistics that further studies within the framework of particular models will be able to make precise \citep[for existing work in this direction, see][]{2002ApJ...565..482G, 2004ApJ...604..775G, 2007ApJ...671..713S, 2007Ap&SS.309..221H}:\\ \\
{\bf $\gamma-$ray beaming fraction.} Outer gap models predict wide $\gamma-$ray beams, with beaming fractions of order unity, while polar cap models predict much smaller beams comparable to those in the radio.
Therefore, outer gap models predict a large ratio of radio-quiet to radio-loud $\gamma-$ray pulsars and a stronger integrated $\gamma-$ray background.
Slot gap models are intermediate.\\ \\
{\bf $\gamma-$ray spectrum.} 
The different classes of models predict different high-energy cutoffs.
Polar cap models, for example, can have hyper-exponential cutoffs that are a signature of $\gamma-B$ pair attenuation of low-altitude emission.
If MSPs are found to contribute a significant fraction of the $\gamma-$ray background, then the spectrum of the latter (possibly after subtracting the blazar component) will inform us about the average MSP spectrum.
This information would be complementary to observations of individual MSPs, since it is conceivable that the brightest MSPs have different properties from the bulk of the population.\\ \\
{\bf Dependence on magnetic and rotational parameters.} Whether MSPs do contribute a significant fraction of the $\gamma-$ray background depends in part on whether the empirically-motivated $L_{\gamma}$ relation in equation (\ref{sqrt Edot prescription}) holds for all pulsars.
Given a MSP population model and if a large fraction of the high-latitude $\gamma-$ray background is found to arise from blazars (or other non-pulsar sources), one can constrain the range of applicability of any particular $\gamma-$ray luminosity law.

\begin{figure*}
\includegraphics[width=1.0\textwidth]{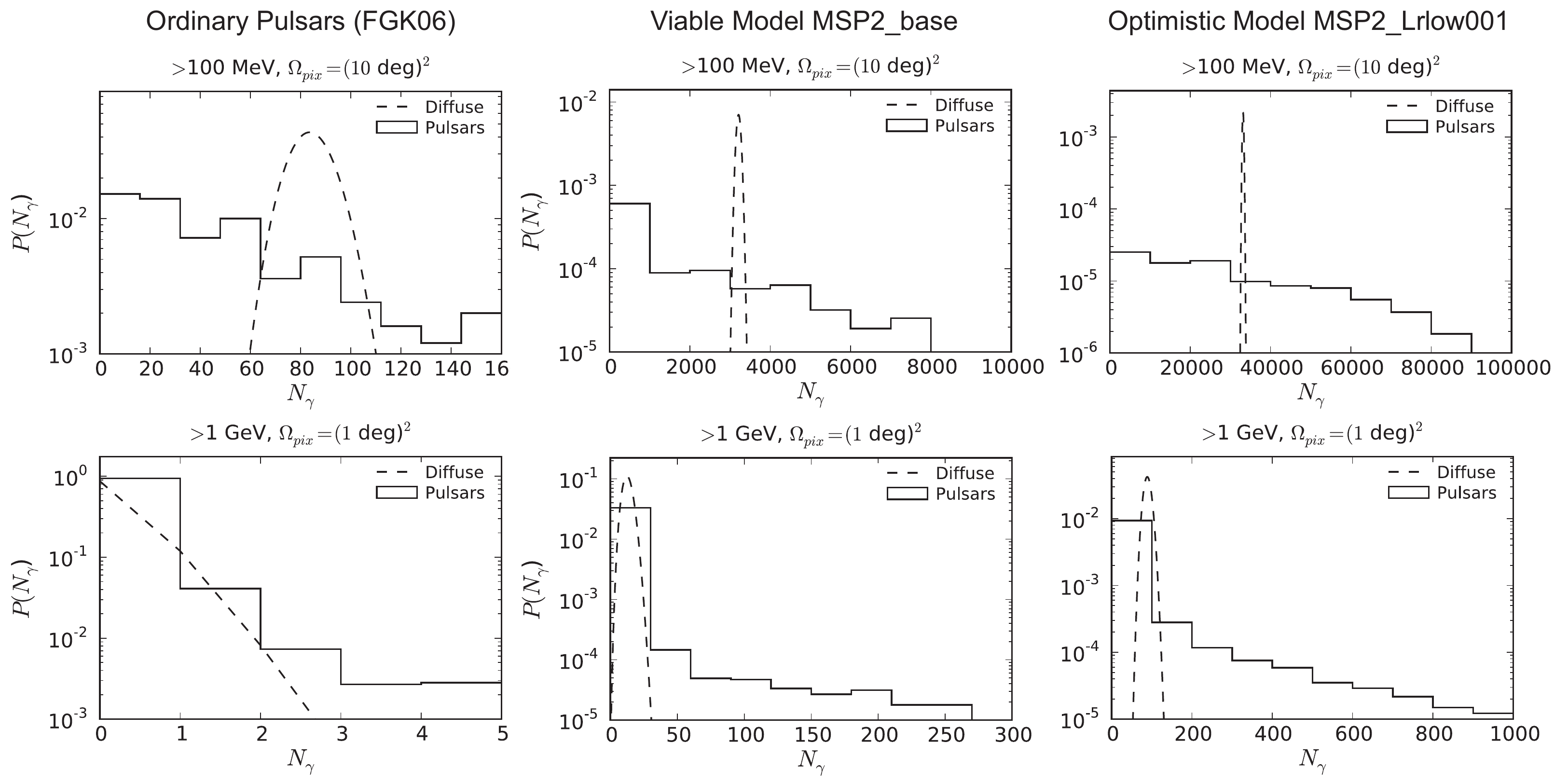}
\caption[]{Illustration of how the pixel count PDF can distinguish between different sources of the $\gamma-$ray background.
We show the normalized PDF for pixels at latitudes $|b|\geq40^{\circ}$ for an exposure $X=2.5\times10^{11}$ cm$^{2}$ s, corresponding to a 5-year all-sky \emph{Fermi} survey.
We consider two pixel size and energy range combinations: $>$100 MeV, $\Omega_{\rm pix}=(10^{\circ})^{2}$ and $>$1 GeV, $\Omega_{\rm pix}=(1^{\circ})^{2}$. 
The left column shows the results for our ordinary pulsar model, the central column the results for a particular MSP model (model {\tt MSP2\_base}) selected for its consistency with existing observations (\S \ref{comparison with existing gamma rays}), and the right column the results for a more optimistic MSP model (model {\tt MSP2\_Lrlow001}).
In each panel, the dashed curve shows the Poisson distribution expected if the background is truly diffuse and uniform on the sky, with the same total number of counts.
Pulsar models are clearly distinct, with much broader PDFs.
The pixel count PDF could also be computed for any particular blazar model and compared with pulsar predictions.}
\label{fluctuations}
\end{figure*} 

\section{Fluctuations in the $\gamma-$ray Sky}
\label{pixel fluctuations}
Most of this work has focused on average $\gamma-$ray background observables, such as its intensity and spectrum.
Fluctuations in the $\gamma-$ray sky can however also be fruitfully exploited to discriminate between different origins of the $\gamma-$ray background, even if its sources are unresolved.
In this section, we show how a particular statistic, the probability distribution function (PDF) of the number of $\gamma-$ray counts in sky pixels, can be used to distinguish the $\gamma-$ray background of pulsars from truly diffuse emission and potentially also from other sources like blazars.\\ \\
We imagine dividing a portion of the sky of interest into $N_{\rm pix}$ angular pixels each of solid angle $\Omega_{\rm pix}$.
The region of the sky is then observed for an exposure $X$ in units of cm$^{2}$ s, reflecting a combination of effective area and integration time.
At the end of the observation period, each pixel has collected a number of $\gamma-$ray photons $N_{\gamma}$.
By combining the data from all the pixels, we can estimate $P(N_{\gamma})$, the PDF of the number of photons collected by an individual pixel.
The method takes advantage of the ability of the telescope to measure the direction of arrival of the $\gamma-$ray photons.
\emph{Fermi}, for instance, has an energy-dependent point spread function, with an angular resolution that improves with increasing energy.
We may thus potentially gain with this approach by considering only photons above a given energy threshold.
The 68\% containment angle for $\emph{Fermi}$ varies from approximately $5^{\circ}$ to $0.5^{\circ}$ from 100 MeV to 1 GeV.\footnote{http://www-glast.slac.stanford.edu/software/IS/ \\ glast\_lat\_performance.htm}

\subsection{Analytical Derivation}
We first assume that the $\gamma-$ray background originates from a population of uncorrelated discrete sources, with a photon flux distribution $dN/dS$ above a prescribed energy threshold, and a total of $N_{\rm src}$ sources in the sky area considered.
By the law of total probability,
\begin{equation}
\label{pixel distribution}
P(N_{\gamma}) = \sum_{k=0}^{\infty} P(N_{\gamma}|~k~{\rm src~in~pix})P(k~{\rm src~in~pix}),
\end{equation}
where `$k$ src in pix' represents the case that exactly $k$ sources lie in the pixel.
The distribution of the number of sources within each pixel is a simple binomial:
\begin{equation}
P(k~{\rm src~in~pix}) = \binom{N_{\rm src}}{k} p_{\rm pix}^{k} (1-p_{\rm pix})^{N_{\rm src}-k},
\end{equation}
where $p_{\rm pix} \equiv N_{\rm pix}^{-1}$ is the probability that a source lies within any particular pixel.
If a pixel contains a single source of photon flux $S$, the number of photons collected after an exposure $X$ has a Poisson distribution with a mean equal to the expected number at the end of the exposure, $SX$.
Taking into account the photon flux distribution,
\begin{equation}
\label{one source case}
P(N_{\gamma}|~1~{\rm src~in~pix}) = 
\frac{1}{N_{\rm src}}
\int_{0}^{\infty} dS
\frac{dN}{dS}
\frac{e^{-SX}(SX)^{N_{\gamma}}}{N_{\gamma}!}.
\end{equation}
Note that because of the integration over the photon flux distribution, this PDF in general differs strongly from a Poisson distribution.\\ \\
If $k>1$ sources lie in the pixel, $N_{\gamma}$ is distributed like the sum of independent random variables with the distribution in equation (\ref{one source case}).
$P(N_{\gamma}|~k~{\rm src~in~pix})$ can then be expressed as a convolution of PDFs of the form of the one-source case.
For small $k$, the PDF will also be non-Poissonian.
For large $k$, however, the central limit theorem implies that the distribution will tend to become Gaussian and therefore lose discriminating power.
A good choice for the pixel size is therefore one for which (given the instrumental limitations) each pixel contains a small number of bright sources, so that $P(N_{\gamma})$ is determined by the small-$k$ $P(N_{\gamma}|~k~{\rm src~in~pix})$ terms in equation (\ref{pixel distribution}).\\ \\
For comparison, consider the case of a $\gamma-$ray background with exactly the same average intensity and spectrum as the discrete case above, but completely diffuse and uniform.
Then $P(N_{\gamma})$ is a simple Poisson distribution with mean equal to the expected number of photons per pixel at the end of the exposure, $\langle N_{\gamma}  \rangle = (N_{\rm src}/N_{\rm pix}) \langle S \rangle X$.
Therefore, even for same intensity and spectrum, the pixel count PDF can distinguish between between a discrete and diffuse origin for the $\gamma-$ray background if the angular resolution of the instrument is sufficient.
This idea can be extended to distinguish competing models that consist of discrete sources.
As equation (\ref{one source case}) shows, $P(N_{\gamma})$ depends on the full photon flux distribution (and total number) of the sources.
The PDF could therefore help distinguish, for example, pulsar and blazar models, or even between different pulsar models.

\subsection{Numerical Models and Application to \emph{Fermi}}
We may use the numerical models of \S \ref{population synthesis models} to compute the pixel count PDF expected from pulsars.
We focus on the $|b|\geq40^{\circ}$ pixels of a 5-year \emph{Fermi} all-sky survey, which corresponds to an exposure of about $X=2.5\times10^{11}$ cm$^{2}$ s \citep[][]{2008JCAP...07..013B}.
We explore two combinations of pixel size and energy range resolvable by \emph{Fermi}: $>$100 MeV, $\Omega_{\rm pix}=(10^{\circ})^{2}$ and $>$1 GeV, $\Omega_{\rm pix}=(1^{\circ})^{2}$.
For these numerical calculations, we divide the sky in our simulations into pixels and simulate an actual observation.
The results are shown in Figure \ref{fluctuations}, with our ordinary pulsar model on the left, one of our viable MSP models (model {\tt MSP2\_base}) at the center, and a more optimistic MSP model (model {\tt MSP2\_Lrlow001}) on the right.
In each panel, the dashed curve shows the Poisson distribution expected if the background is truly diffuse and uniform on the sky (as may be expected, for example, in some dark matter annihilation scenarios), with the same total number of counts.
Pulsar models are clearly distinct, with much broader PDFs.
The pixel count PDF could also be computed for any particular blazar model and compared with pulsar predictions to help identify which are the dominant sources.
Note, however, that different population models may yield different PDFs and so a proper analysis of the data will need to explore a range of both pulsar and blazar models in order to robustly distinguish between the two hypotheses.

\section{Conclusions}
\label{conclusion}
We have estimated the contribution of Galactic pulsars to the diffuse $\gamma-$ray background, focusing on the high-latitude part ($|b|\geq 40^{\circ}$) that could be confused with an extragalactic component.
We have considered both the population of ordinary pulsars and the population of recycled MSPs.
For the ordinary pulsars, the underlying population parameters were constrained in detail by the radio population study of FGK06 and taken as fixed.
Since the MSP population is not as well constrained observationally due to more severe and complex selection effects, we instead explored a range of models with fiducial parameters based on the literature, also calibrated to the number of detections in radio surveys. 
We examined how the predicted $\gamma-$ray background, as well as the predicted number early \emph{Fermi} detections, depend on the population parameters as a guide to how $\gamma-$ray measurements can usefully constrain the MSP population.
To assign $\gamma-$ray luminosities to synthetic pulsars, we adopted a simple empirical prescription motivated by \emph{EGRET} results and which also provides a good fit to early \emph{Fermi} and \emph{AGILE} observations, including a some MSPs.
According to this prescription (eq. \ref{sqrt Edot prescription}) that upcoming \emph{Fermi} observations will help refine, the $\gamma-$ray luminosity of each pulsar is a simple function of its period and period derivative, up to a maximum fraction of the total power available from spin-down.\\ \\
We have found that while ordinary pulsars are expected to contribute a relatively small fraction $\sim10^{-3}$ of the high-latitude $\gamma-$ray background, MSPs could in principle easily contribute a significant fraction of the measured high-latitude intensity, $I_{X}\sim1\times10^{-5}$ ph s$^{-1}$ cm$^{-2}$ sr$^{-1}$.
It is therefore possible to use measurements of the $\gamma-$ray background to usefully constrain this population (\S \ref{comparison with existing gamma rays}); many population models are in fact already ruled out by existing $\gamma-$ray observations.
Models that are consistent with early \emph{Fermi} results suggest that MSPs could contribute $\sim5-15\%$ of the high-latitude $\gamma-$ray background if the measurements obtained by subtracting a cosmic ray model are correct, but up to $\sim 20$\% or more if the more model-independent analysis of \cite{2004JCAP...04..006K} is more accurate.
This conclusion is actually independent of the assumed details of the MSP population and rests principally on the simple fact that the ages of MSPs generally exceed the oscillation time across the plane of the Galaxy ($\sim100$ Myr) by a large factor.
Unlike the $\gamma-$ray bright ordinary pulsars which are concentrated near the plane, the $\gamma-$ray bright MSPs are equally prevalent at all latitudes (modulo the population scale height).
This results in a $\log{N}-\log{S}$ flux distribution that is nearly Euclidean at the high-end for MSPs, which can be used to relate the integrated background to the number of resolved point sources above a given flux threshold.\\ \\
For our adopted $\gamma-$ray prescription, the $\gamma-$ray background intensity is particularly sensitive to the period of the most rapidly-rotating MSPs.
If this maximum spin rate can be constrained to be significantly below the break-up upper limit, it would provide information on the processes that limit the spin frequency.
This would constrain the conditions of spin-magnetic equilibrium in which MSPs are thought to acquire their high rotation frequencies, and may be indicative of braking by gravitational wave emission, which would have important ramifications for direct detection efforts by \emph{LIGO}.\footnote{http://www.ligo.caltech.edu}\\ \\
Given that pulsars, and MSPs in particular, appear quite capable of contributing significantly to the $\gamma-$ray background, it will clearly be worthwhile for upcoming $\emph{Fermi}$ studies to check and refine the $\gamma-$ray prescription we have used in our calculations.
More detailed population synthesis studies of MSPs in the radio, for which there has been relatively little work, will also be very useful.
If the $\gamma-$ray contribution of Galactic MSPs proves to be important, it will be necessary to also subtract it in order to obtain an accurate measurement of the extragalactic background.
MSPs will have to be considered when attempting to detect a putative component from dark matter annihilation, which is likely to produce a subdominant signal.
As we showed in \S \ref{pixel fluctuations}, fluctuations in the $\gamma-$sky (which depend on the full $\log{N}-\log{S}$ flux distribution) can be fruitfully exploited to distinguish between different sources of the $\gamma-$ray background.
As the telescope exposure increases, more of the unresolved discrete sources contributing to the fluctuations will be individually detected. 
\\ \\
As a final recommendation for future observations, we emphasize the importance of blind searches since a substantial fraction of the $\gamma-$ray background could originate from pulsars that are not visible in the radio.
This is especially true for MSPs that can be difficult to detect in the radio (either because of instrumental limitations, dispersion or scattering by intervening interstellar electrons, or because of orbital acceleration or eclipses in binary systems) but may be more prevalent as $\gamma-$ray sources.
At present, blind time-differencing searches for periodicity by \emph{Fermi} are limited to $P>16$ ms and $\dot{P}/P<1.25\times10^{-11}$ s$^{-1}$ \citep[][]{2009arXiv0901.3405S}.
If possible (given the limitations imposed by the small numbers of counts) the \emph{Fermi} search space should be increased to allow the detection of shorter-period MSPs and also for acceleration in binary systems, as these should be both numerous and may be significant contributors to the diffuse background.

\section*{Acknowledgments}
The authors are grateful to Vicky Kaspi for useful discussions and detailed comments on an earlier version of the manuscript.
CAFG is supported by an NSERC Post-graduate Fellowship and a supplement from the Canadian Space Agency.

\appendix
\section{RELATION BETWEEN RESOLVED SOURCES AND THE INTEGRATED BACKGROUND}
\label{relation between resolved sources and the background}
The contribution of a source population to the integrated background intensity, $I$, is in general not independent of the number of resolved sources above a given flux threshold, $N(>S_{\rm th})$.
If the $\log{N}-\log{S}$ flux distribution is known, the two are in fact a direct function of one another.
While the actual full flux distribution of the Galactic pulsar population is not yet empirically measured, we can combine the theoretical predictions of \S \ref{population synthesis models} with early \emph{Fermi} results to more model-independently quantify the contribution of pulsars to the $\gamma-$ray background intensity.\\ \\
For simplicity, let us assume that the pulsar $\gamma-$ray flux distribution has a simple power-law form above a lower limit $S_{\rm min}$,
\begin{equation}
\label{dN dS function}
\frac{dN}{dS} =  \left\{
\begin{array}{rl}
A S^{-\beta} & \text{if } S \geq S_{\rm min}\\
0 & \text{otherwise}
\end{array} \right. .
\end{equation}
Then, provided that $S_{\rm th}>S_{\rm min}$ and $\beta>2$,
\begin{equation}
N(>S_{\rm th}) = \int_{S_{\rm th}}^{\infty} dN =  \frac{A}{\beta-1} S_{\rm th}^{1-\beta},
\end{equation}
while the integrated flux from all the sources
\begin{equation}
S_{\rm tot} = \int_{0}^{\infty} dN S = \frac{A}{\beta - 2} S_{\rm min}^{2-\beta}.
\end{equation}
Defining the average background intensity over a solid angle $\Omega$ as $I_{\rm psr} \equiv S_{\rm tot} / \Omega$, we obtain the simple relation
\begin{equation}
I_{\rm psr} = \frac{S_{\rm min}}{\Omega}
\left( \frac{\beta-1}{\beta-2} \right)
\left( \frac{S_{\rm min}}{S_{\rm th}} \right)^{1-\beta} N(>S_{\rm th}).
\end{equation}
To compare with existing observations, we can consider the portion of the sky at Galactic latitudes $|b|\geq b_{\rm min}$, for which $\Omega = 4 \pi (1 - \sin{b_{\rm min}})$. 
We base our discussion on the analysis of the $|b|\geq10^{\circ}$ sources detected by \emph{Fermi} in its first three months of operation reported by \cite{2009ApJ...700..597A} (see also \S \ref{which models pass tests}).
This sample is complete for fluxes $S>1.25\times10^{-7}$ ph s$^{-1}$ cm$^{-2}$ at energies $>$100 MeV.
Among the sources with flux above this threshold, two are new pulsars discovered by \emph{Fermi} (LAT PSR J1836+5925 and LAT PSR J0007+7303)
and two more are unidentified.
While 40 are associated at high-confidence with AGNs, the high-confidence associations are only defined as having probability $>$90\% of being genuine, so that a few could be spurious.
In addition, PSR J0357.5+3205 is a previously-known MSP with $\gamma-$ray flux just below the completeness threshold, but consistent with a value above the threshold within the uncertainty.
Because the $\gamma-$ray beams of pulsars appear generally wider than their radio beams and because many pulsars have not yet been detected at other wavelengths, we can expect at least as many $\gamma-$ray pulsars to be unidentified as sources with known counterparts.
It is thus likely that the sources in the flux limited sample that are not associated with a known AGN are pulsars.
Conservatively and accounting for the fact that some of the potential sources may not be MSPs, we fiducially assume that there are four $\gamma-$ray MSPs in the flux-limited sample.
For an Euclidean source distribution ($\beta=2.5$; see Appendix \ref{slope of the flux distribution} below), the \emph{Fermi} resolved sources thus allow for a contribution of pulsars to the $|b|\geq10^{\circ}$ $\gamma-$ray background at energies $>100$ MeV of
\begin{align}
\label{bkg analytic eq}
I_{\rm psr}(|b|\geq10^{\circ}) \sim &~0.5 \times 10^{-5}{\rm~ph~s^{-1}~cm^{-2}~sr^{-1}} \notag \\
& \times \left( \frac{S_{\rm min}}{10^{-10}~{\rm ph~s^{-1}~cm^{-2}}} \right)^{-0.5} \notag \\
& \times \left( \frac{S_{\rm th}}{1.25\times10^{-7}~{\rm ph~s^{-1}~cm^{-2}}} \right)^{1.5} \notag \\
& \times \left( \frac{N(>S_{\rm th})}{4} \right),
\end{align}
with a 50\% uncertainty from small-number statistics only.
In the above estimate, the $S_{\min}$ factor was approximately set to the turn over of the MSP $\log{N}-\log{S}$ distribution seen in our MSP population synthesis models (\S \ref{population synthesis models}).
If the \emph{Fermi} flux-limited sample contains spurious associations of sources with AGNs because of positional coincidence, then the true $N(>S_{\rm th})$ for MSPs could be higher and we would therefore expect a larger contribution from these sources to the background intensity.
\\ \\
In this work, as in the study of \cite{2004JCAP...04..006K}, we defined the intensity of the high-latitude $\gamma-$ray background as the $|b|\geq40^{\circ}$ average (\S \ref{model results}).
However, the \emph{Fermi} flux limited sample on which we base our point source comparison covers $|b|\geq10^{\circ}$ \citep{2009ApJ...700..597A}.
Assuming that the latitude profile is disc-like ($\sim 1/ \sin{|b|}$; Figure \ref{maps}), there is a simple expression for the ratio of the average intensities at $|b|\geq b_{\rm min}$ for different values of $b_{\rm min}$.
This expression follows from the result $\langle (\sin{|b|})^{-1} \rangle (|b|\geq b_{\rm min})=\ln{[(\sin{|b_{\rm min}|})^{-1}]}$ for the solid angle average:
\begin{equation}
\frac{I(|b|\geq b_{\rm min,1})}{I(|b|\geq b_{\rm min,2})} = 
\frac{\ln{[(\sin{|b_{\rm min,1}|})^{-1}]}}{\ln{[(\sin{|b_{\rm min,2}|})^{-1}]}}.
\end{equation}
Numerically, $I(|b|\geq 40^{\circ})\approx 0.25 I(|b|\geq 10^{\circ})$ and the numerical factor in the analytic estimate of equation (\ref{bkg analytic eq}) becomes $0.125 \times 10^{-5}{\rm~ph~s^{-1}~cm^{-2}~sr^{-1}}$ for $I_{\rm psr}(|b|\geq40^{\circ})$.

\section{SLOPE OF THE FLUX DISTRIBUTION}
\label{slope of the flux distribution}
In order to relate the number of resolved point sources to the integrated flux from a population as in the previous section, it is necessary to know the slope $\beta$ for the $dN/dS$ flux distribution (eq. (\ref{dN dS function})).
In Figure \ref{maps}, we plot slopes in the $\log{N}-\log{S}$ panels for the ordinary pulsar and MSP models to guide the eye; based on these, we approximated the high-flux end of the MSP distribution to have an Euclidean slope $\beta=2.5$ in our analytical estimates.
In this section, we show how the different slopes arise and argue that they are robust to the details of our population models.\\ \\
Let us first assume that all the sources have an intrinsic luminosity $L$.
We consider three idealized cases, corresponding to homogeneous spatial distributions along a one-dimension line (linear), in a two-dimensional plane (planar), and in three-dimensional space (Euclidean).
In general,
\begin{equation}
\frac{dN}{dS}
=
\frac{dN}{dr}
\left|
\frac{dr}{dS}
\right|,
\end{equation}
where $r$ is the distance from the observer.
Since $S = L / 4 \pi r^{2}$, $|dr/dS|=2\pi r^{3}/L$ for all geometries.
If we denote by $\lambda$, $\sigma$, and $\rho$ the linear, planar, and Euclidean spatial source densities we have
\begin{equation}
\frac{dN}{dr} =  \left\{
\begin{array}{cl}
\lambda & \text{if linear}\\
2\pi r \sigma & \text{if planar}\\
4\pi r^{2} \rho & \text{if Euclidean}\\
\end{array} \right. .
\end{equation}
Factoring out the constants, this gives
\begin{equation}
\frac{dN}{dS}(L) =  \left\{
\begin{array}{cl}
f(L) S^{-3/2} & \text{if linear}\\
g(L) S^{-2} & \text{if planar}\\
h(L) S^{-5/2} & \text{if Euclidean}\\
\end{array} \right.
\end{equation}
for $f$, $g$, and $h$ that are functions of $L$ and where we have written $(dN/dS)(L)$ to emphasize we have assumed that all the sources have the same luminosity.\\ \\
Consider now the case of an arbitrary luminosity function $\phi(L)$, normalized such that $\int dL \phi(L)=1$.
Then the total flux distribution is simply given by
\begin{equation}
\frac{dN}{dS} = \int dL \phi(L) \frac{dN}{dS}(L).
\end{equation}
Explicitly,
\begin{equation}
\frac{dN}{dS} =  \left\{
\begin{array}{cl}
( \int dL f(L) ) S^{-3/2} & \text{if linear}\\
( \int dL g(L) ) S^{-2} & \text{if planar}\\
( \int dL h(L) ) S^{-5/2} & \text{if Euclidean}\\
\end{array} \right. ,
\end{equation}
which shows that the slope of the distribution is preserved regardless of the luminosity function.
The $dN/dS$ slope however depends on the geometrical distribution of the sources.
In cases in which the spatial source distribution is bounded, for example, one can show that the luminosity function does shape the flux distribution to some extent.
Empirically, we find that the high-flux end of the MSP $\log{N}-\log{S}$ distribution has a nearly Euclidean slope $\beta=2.5$ (Figure \ref{maps}) regardless of the details of the population parameters (which do affect the luminosity function).
This reflects the near-Euclidean nature of the spatial distribution in the Solar neighborhood.
The distribution departs from this slope at lower fluxes principally owing to the finite scale height of the distribution above the Galactic plane.
On the other hand, the high-flux end of the ordinary pulsar population is intermediate between the planar ($\beta=2$) and linear ($\beta=3/2$) cases, reflecting the fact that the brightest $\gamma-$ray pulsars are the young ones located in the Galactic plane and concentrated in spiral arms. 

\bibliography{references} 

\onecolumn

\begin{landscape}

\setcounter{table}{1}
\setcounter{section}{0}

\begin{longtable}{p{1.0in}ccccccccc}

\caption{Pulsar Population Models}\label{msp models table} \\

\hline \hline \\[-2ex]
 & 
 $\sigma_{r}$ & 
 $\langle z \rangle$ & 
 ($P_{\rm low},~P_{\rm break},$ &
 $(\langle \log{B} \rangle,~\sigma_{\log{B}})$ & 
 $(L_{\rm r,low},~L_{\rm r,break},$ &
 $\mathscr{N}_{\rm MSP}$ &
$N_{\rm comp}(|b|\geq10^{\circ})$ & 
 $I_{\rm psr}(|b|\geq40^{\circ})$ \\[-0.25ex]  
\endfirsthead

\multicolumn{9}{c}{{\tablename} \thetable{} -- Continued} \\[0.5ex]  
  \\[-1.8ex]
\endhead

\\
\endfoot

\hline
\\[0.2ex]  \multicolumn{9}{p{7in}}{Lengths are in kpc, periods in s, magnetic fields in G, radio luminosities in mJy kpc$^{2}$ at 1.4 GHz, and the $\gamma-$ray background intensity is in 10$^{-5}$ ph s$^{-1}$ cm$^{-2}$ sr$^{-1}$.
Asterisks indicate that a single power-law index is assumed throughout, with no break.}   \\[2.8ex] 
\endlastfoot


 &  &  & $~~~~~m_{P,1},~m_{P,2})$ &  & $~~~~~m_{L,1},~m_{L,2})$ &  &  &  \\[+2pt]
\hline
Ordinary FGK06 &  *            & *                   & *           & *           & * & * & 0 & $8.6\times10^{-4}$ \\[+2pt]
\hline
{\tt MSP1\_base} &  5            & 1                   & (1.0,*,-1,*)          & $(8.5,~0.2)$ & (0.1,*,-1,*)            & 0.0317 & 977 & 4.05 \\[+2pt]
\hline
{\tt MSP1\_Lrlow001} &  5             & 1                   & (1.0,*,-1,*)          & $(8.5,~0.2)$ & (0.01,*,-1,*)          & 0.312 & 9971 & 42.5 \\[+2pt]
{\tt MSP1\_Lrlow10} &  5             & 1                   & (1.0,*,-1,*)          & $(8.5,~0.2)$ & (1.0,*,-1,*)           & 0.00392 & 133 & 0.539 \\[+2pt]
\hline
{\tt MSP1\_sigr3} &  3            & 1                   & (1.0,*,-1,*)          & $(8.5,~0.2)$ & (0.1,*,-1,*)            & 0.0462 & 904 & 1.54 \\[+2pt]
{\tt MSP1\_sigr7} &  7            & 1                   & (1.0,*,-1,*)          & $(8.5,~0.2)$ & (0.1,*,-1,*)            & 0.0382 & 1178 & 5.53 \\[+2pt]
\hline
{\tt MSP1\_z05} &  5            & 0.5                 & (1.0,*,-1,*)          & $(8.5,~0.2)$ & (0.1,*,-1,*)            & 0.0252 & 529 & 2.67 \\[+2pt]
{\tt MSP1\_z15} &  5           & 1.5                 & (1.0,*,-1,*)          & $(8.5,~0.2)$ & (0.1,*,-1,*)            & 0.0410 & 1407 & 5.95 \\[+2pt]
{\tt MSP1\_z20} &  5           & 2.0                 & (1.0,*,-1,*)          & $(8.5,~0.2)$ & (0.1,*,-1,*)            & 0.0472 & 1538 & 7.38 \\[+2pt]
\hline
{\tt MSP1\_Plow065} &  5           & 1                   & (0.65,*,-1,*)          & $(8.5,~0.2)$ & (0.1,*,-1,*)           & 0.0438 & 4687 & 23.8 \\[+2pt]
{\tt MSP1\_Plow125} &  5           & 1                   & (1.25,*,-1,*)           & $(8.5,~0.2)$ & (0.1,*,-1,*)          & 0.0274 & 381 & 1.54 \\[+2pt]
{\tt MSP1\_Plow15} &  5           & 1                   & (1.5,*,-1,*)           & $(8.5,~0.2)$ & (0.1,*,-1,*)           & 0.0247 & 164 & 0.693 \\[+2pt]
\hline
{\tt MSP1\_Pbreak15} &  5            & 1                   & (1.0,1.5,0,-1)          & $(8.5,~0.2)$ & (0.1,*,-1,*)            & 0.0295 & 823 & 3.76 \\[+2pt]
{\tt MSP1\_Pbreak3} &  5            & 1                   & (1.0,3,0,-1)          & $(8.5,~0.2)$ & (0.1,*,-1,*)            & 0.0259 &  500 & 2.08 \\[+2pt]
{\tt MSP1\_Pbreak5} &  5            & 1                   & (1.0,5,0,-1)          & $(8.5,~0.2)$ & (0.1,*,-1,*)            & 0.0228 &  342 & 1.51 \\[+2pt]
{\tt MSP1\_Pbreak10} &  5            & 1                   & (1.0,10,0,-1)          & $(8.5,~0.2)$ & (0.1,*,-1,*)            & 0.0182 & 203 & 0.938 \\[+2pt]
\hline
{\tt MSP1\_logB8} &  5            & 1                   & (1.0,*,-1,*)          & $(8,~0.2)$ & (0.1,*,-1,*)            & 0.0347 & 84 & 0.602 \\[+2pt]
{\tt MSP1\_logB825} &  5            & 1                   & (1.0,*,-1,*)          & $(8.25,~0.2)$ & (0.1,*,-1,*)            & 0.0326 & 367 & 1.54 \\[+2pt]
{\tt MSP1\_logB875} &  5            & 1                   & (1.0,*,-1,*)          & $(8.75,~0.2)$ & (0.1,*,-1,*)            & 0.0313 & 2320 & 11.70 \\[+2pt]
\hline
{\tt MSP1\_sigB03} &  5            & 1                   & (1.0,*,-1,*)          & $(8.5,~0.3)$ & (0.1,*,-1,*)            & 0.0317 & 1275 & 6.21 \\[+2pt]
{\tt MSP1\_sigB04} &  5            & 1                   & (1.0,*,-1,*)          & $(8.5,~0.4)$ & (0.1,*,-1,*)            & 0.0325 & 1605 & 9.31 \\[+2pt]
{\tt MSP1\_sigB05} &  5            & 1                   & (1.0,*,-1,*)          & $(8.5,~0.5)$ & (0.1,*,-1,*)            & 0.0314 & 2148 & 12.4 \\[+2pt]
\hline
{\tt MSP2\_base} &  5            & 1                   & (1.5,*,-1,*)          & $(8,~0.2)$ & (0.1,*,-1,*)            & 0.0245  & 4 & 0.0775 \\[+2pt]
\hline
{\tt MSP2\_Lrlow0005} &  5            & 1                   & (1.5,*,-1,*)          & $(8,~0.2)$ & (0.005,*,-1,*)            & 0.495 & 195 & 1.51 \\[+2pt]
{\tt MSP2\_Lrlow001} &  5            & 1                   & (1.5,*,-1,*)          & $(8,~0.2)$ & (0.01,*,-1,*)            & 0.214 & 79 & 0.698 \\[+2pt]
{\tt MSP2\_Lrlow002} &  5            & 1                   & (1.5,*,-1,*)          & $(8,~0.2)$ & (0.02,*,-1,*)            & 0.115 & 44 & 0.438 \\[+2pt]
{\tt MSP2\_Lrlow005} &  5            & 1                   & (1.5,*,-1,*)          & $(8,~0.2)$ & (0.05,*,-1,*)            & 0.0459 & 13 & 0.175 \\[+2pt]

\end{longtable}

\end{landscape}

\end{document}